\documentclass[12pt,preprint]{aastex}

\slugcomment{\bf Accepted August 8, 2003 to AJ}

\shorttitle{Stellar Properties of Pre--Main Sequence Stars from High Resolution Near--IR Spectra}
\shortauthors{Doppmann et al.}

\begin{document}

\title{Stellar Properties of Pre--Main Sequence Stars from High Resolution Near--IR Spectra}

\author {G. W. Doppmann,\altaffilmark{1,2,3} D.T. Jaffe,\altaffilmark{2,3} R.J. White\altaffilmark{3,4}}
\altaffiltext{1}{NASA Ames Research Center, MS 245-6, Moffett Field, CA  94035-1000}
\altaffiltext{2}{Visiting Astronomer, Kitt Peak National Observatory, National Optical Astronomy Observatory, which is operated by the Association of Universities for Research in Astronomy, Inc. (AURA) under cooperative agreement with the National Science Foundation.}
\altaffiltext{3}{Department of Astronomy, 1 University Station C1400, Austin, TX 78712-1083}
\altaffiltext{4}{California Institute of Technology, Department of Astronomy, MS 105-24, Pasadena, CA  91125}

\email{gdoppmann@mail.arc.nasa.gov}
\email{dtj@astro.as.utexas.edu}
\email{rjw@astro.caltech.edu}

\begin{abstract}
  
  We present high resolution (R=50,000) spectra at 2.2 $\mu$m of 16
  young stars in the $\rho$~Ophiuchi dark cloud.  Photospheric
  features are detected in the spectra of 11 of these sources, all
  Class II young stellar objects.  The 5 featureless spectra consist
  of 2 Class I, 2 Class I.5, \& 1 Class II.  One star, GSS 29, is
  identified as a spectroscopic binary based on radial velocity
  variations.  The radial velocities for the remaining sample are
  consistent with $^{12}$CO and H$_2$CO gas velocities and further
  confirm the membership of the sources in the $\rho$~Ophiuchi
  cluster.  For the 10 spectroscopically single Class II sources, we
  measure effective temperatures, continuum veiling, and $v\sin i$
  rotation from the shapes and strengths of atomic photospheric lines
  by comparing to spectral synthesis models at 2.2 $\mu$m.  We measure
  surface gravities in 2 stars from the integrated line flux ratio of
  the $^{12}$CO line region at 2.3 $\mu$m and the Na I line region at
  2.2 $\mu$m.  Although the majority (8/10) of the Class II stars have
  similar effective temperatures (3530 K $\pm$ 100 K), they exhibit a
  large spread in bolometric luminosities (factor $\sim$8), as derived
  from near--IR photometry.  In the two stars where we have surface
  gravity measurements from spectroscopy, the photometrically derived
  luminosities are systematically higher than the spectroscopic
  luminosities.  The spread in the photometrically derived
  luminosities in our other sources suggests either a large spread in
  stellar ages, or non-photospheric emission in the J--band since
  anomalous and significant veiling at J has been observed in other T
  Tauri stars.  Our spectroscopic luminosities result in older ages on
  the Hertzsprung--Russell diagram than is suggested by photometry at
  J or K.  Most of our sources show a larger amount of continuum
  excess (F$_{\rm Kex}$) than stellar flux at 2.2 $\mu$m (F$_{\rm
  K*}$), substantially higher in many cases (r$_{\rm K}$~$\equiv$~
  F$_{\rm Kex}$/F$_{\rm K*}$~=~0.3--4.5).  The derived veiling values
  at K (r$_{\rm K}$) appear correlated with mid--IR disk luminosity,
  and with Brackett $\gamma$ equivalent width, corrected for veiling.
  The derived $v\sin i$ rotation is substantial (12--39 km s$^{-1}$),
  but systematically less than the rotation measured in Class I.5
  (flat) and Class I sources from other studies in Ophiuchus.  In four
  stars (Class I and I.5 sources), the absence of any photospheric
  lines is likely due to large continuum excess and/or rapid rotation
  if the stars have late--type photospheres.

\end{abstract}


\section{Introduction}

High resolution spectroscopy can be a valuable tool for extracting
stellar properties of highly extincted young stellar objects (YSOs).
Spectroscopy offers a more direct method than photometry for measuring
properties of obscured young stars.  Absorption line equivalent widths
and ratios are not dependent on extinction.  The intrinsic shapes of
photospheric absorption lines, which can be fully resolved in cool
stars (T $<$ 5500 K) by high spectral resolution (R $\equiv$
$\lambda$/$\Delta\lambda$ $\geq$ 10,000--20,000) observations, contain
stellar kinematic and temperature information.  Observations of
photospheric lines in obscured stars at high spectral resolution
permit us to measure (1) precise radial velocities from line shifts,
(2) $v\sin i$ rotational velocities from line widths, (3) effective
temperatures and surface gravities from line shapes and line ratios,
and (4) the continuum ``veiling'' by hot dust from the line depths.

Unfortunately, it is difficult to obtain YSO spectra at optical
wavelengths because of the heavy extinction in the cores of star
forming regions, such as in the $\rho$~Ophiuchi dark cloud
\citep{wilking1983}.  In addition, large continuum excesses from
shocks due to accretion of disk material onto the YSO surface often
overwhelm the stellar flux in the optical and UV
\citep{gullbring2000}.  In the mid--IR, the thermal emission
from the surrounding disk can also substantially exceed the flux from
the photosphere \citep{fazio1976,lada1984,bontemps2001}.  The standard
model of a young star surrounded by a circumstellar disk suggests that
the photospheric emission should dominate the observed emission at
near--IR wavelengths.  \citep{strom1988,kenyon1990}.
Consequently, the near--IR is the best wavelength region to observe
photospheric flux from YSOs and can serve as a window through which to
observe their physical properties and thereby to study earlier stages
of star formation and disk accretion than are possible at optical
wavelengths.

Clusters are important to study since they represent the dominant mode
of star formation \citep{miller1978,shu1987,lada2003}.  Since stars form from
the collapse of dense cores within molecular clouds
\citep{hartmann1997,evans1999}, the young stars we observe in a
cluster have formed from the same material within the parent molecular
cloud at roughly the same time.  Therefore, young clusters present a
homogeneous sample of stars at similar \citep[though not necessarily
equal,][]{lada2003} ages.  Embedded clusters are of particular
interest because they potentially contain the youngest observable
stars \citep{wilking1983, lada1991}.

One of the nearest examples of a young star forming cluster is the
dark cloud in the $\rho$~Ophiuchus region \citep[D = 145
pc,][]{dezeeuw1999}.  The compact (d $<$ 1 pc) core of the cloud has
been well--studied because it contains a large number of embedded low
mass stars.  The first infrared surveys of the Ophiuchus complex
resulted in the discovery of dozens of embedded YSOs within the more
extincted regions of the cloud \citep{grasdalen1973, vrba1975,
elias1978a, elias1978b}.  \citet{wilking1983}, surveyed a 10$\arcmin $
$\times$ 10.5$\arcmin$ region within the A$_{\rm v}$ = 50 boundary
(derived from their C$^{18}$O contour map) in the $\rho$~Ophiuchi
cloud core.  With the development of infrared detector arrays, large
surveys have been carried out to verify cluster membership, detect the
presence of disks, and make estimates of the star formation efficiency
\citep{greene1992, comeron1993, strom1995}.  Mid--IR photometry has
been employed to characterize the circumstellar disks around pre--main
sequence (PMS) stars in this cloud \citep{GWAYL1994, bontemps2001}.  A
magnitude limited (K $\leq$ 14.0) near--IR imaging survey of a 2.2 pc$^2$
region of $\rho$~Oph was carried out by \citet{barsony1997}.

In the present study, we will build on the earlier work by using high
resolution spectra of a sample of PMS stars in $\rho$~Oph to measure
the physical characteristics important to the PMS evolutionary
stage. High resolution spectroscopy permits us to make direct
measurements of T$_{\rm eff}$, $\log g$, rotation, radial velocity,
and excess (non--photospheric) emission.  By comparing the effective
temperature and gravity or luminosity to evolutionary model tracks on
the Hertzsprung--Russell (H--R) diagram, we can obtain estimates for
YSO masses and ages in a sample that is not limited to binary systems.
Linking the physical characteristics of young stars (singles and
binaries, alike) to their PMS evolutionary stage is therefore a
primary goal.  The $v\sin i$ rotation of PMS stars will give us
insights into the relationship between the young stars and their
circumstellar disks.  Quantifying the amount of infrared continuum
excess allows us to probe the conditions in the disk.  Although high
resolution spectroscopic observations have revealed some of these
properties for a few stars in $\rho$~Ophiuchi \citep[e.g.][]
{bouvier1986,casali1996,greene1997,greene2000,greene2002}, no
systematic survey has yet been completed.

In $\S$ 2, we describe how we obtained our observations and chose our
sample in Ophiuchus.  We present the results of our spectral fitting
technique \citep[][hereafter DJ03]{doppmann2002a} applied to the
PHOENIX spectra of our sample of Ophiuchus YSOs, in $\S$ 3.  In $\S$ 4
and $\S$ 5, we discuss the significance of these results, compare
luminosities derived from photometry and spectroscopy, and put our
results in the context of previous work.  We summarize our conclusions
in $\S$ 6.

\section{Observations and Data Reduction}

\subsection{Selection of the Sample}

The observational goal of this study was to observe a sample of YSOs
within the A$_{\rm v}$~=~50 boundary in the $\rho$~Ophiuchi cloud core
\citep{wilking1983}.  We imposed the following selection criteria to
determine our sample: (1) We selected only PMS stars that were labeled
as Class I, I.5, or II by \citet{luhman1999}(hereafter LR99) or
\citet{greene1996} based on the 2.2 -- 10 $\mu$m spectral index using
photometry of \citet{GWAYL1994}.  (2) We required a signal--to--noise
ratio (S/N) $>$ 20 within one hour to obtain high resolution spectra
(R $\cong$ 50,000).  This resulted in a sample of 18 YSOs with m$_{\rm
K} \leq 8.4$, listed in Table~\ref{tbl-oph.sample}.

\subsection{Spectroscopic Observations}

From the sample of 18 young stars in Ophiuchus in Table 1, we obtained
near--IR spectra of 16 in May 2000 with the PHOENIX spectrograph on the
Kitt Peak 4 meter \citep{hinkle1998}.  Two sources (GY 168 and GY 110)
in the magnitude--limited sample were excluded because there were no
nearby optical stars required for guiding on the KPNO 4 meter.  The
positions and alternate source names for the Ophiuchus targets can be
found in \citet{barsony1997}. Spectra were imaged on an Aladdin II
InSb detector, 1024 $\times$ 512 (used) with 27 $\mu$m pixels.  We
chose the spectral region at 2.21 $\mu$m to measure stellar features
in low mass stars since this wavelength region has strong photospheric
lines in late type stars whose shapes and depths are sensitive to
changes in temperature \citep[DJ03]{kleinmann1986}.

Centered at 2.207 $\mu$m, we obtained 0.0095 $\mu$m of spectral
coverage (the ``Na interval'', hereafter).  With a 1$\arcsec$ (4
pixel) slit we achieved a resolving power of R=50,000.  15$\arcsec$
nods east--west along the slit allowed sequential spectra to be taken
in two beam positions on the array for removal of foreground sky
emission.  Targets were kept in the slit by guiding on optical stars
within the 5$\arcmin$ field of the CCD off--axis guider.  Single
on--chip integrations with PHOENIX were 5--25 minutes, resulting in a
S/N $\sim$ 40--70 obtained in a single beam subtracted pair.  We
estimated the S/N in our spectra from variations in the
photospherically featureless continuum shortward of 2.206 $\mu$m.  Hot
stars were observed at the same airmass throughout each night for
telluric line cancellation.  Seven spectral type standards were also
observed for comparison in our spectroscopic analysis ($\S$
\ref{sect-spect.analysis}).

Supplementary high resolution spectra (R $\sim$ 21,000) of the (2--0)
$^{12}$CO bandhead (2.2935 $\mu$m) of two sources (IRS 2 and GY 314)
were generously provided by Tom Greene, fully reduced from previously
published results \citep{greene1997} using the CSHELL spectrograph
\citep{greene1993} on the IRTF 3 meter.  The combination of these two
data sets allowed us to make measurements of surface gravities using
the combined spectral windows (see $\S$ \ref{sect-spec.logg}).

\subsection{Data Reduction}

All data were reduced using IRAF.  Source frames at different slit
positions were differenced, and then divided by flat fields.  The flat
field source for PHOENIX was a continuum lamp that fully illuminated
the slit.  Bad pixels were removed by interpolation.  All spectra were
optimally extracted using the {\tt apall} package in IRAF.  Telluric
absorption lines (H$_2$O \& CH$_4$) were employed for wavelength
calibration using the HITRAN molecular absorption database
\citep{rothman1992}, and were removed by division of spectra from hot,
early--type stars at the same airmass (within 5$^\circ$ of the target)
at each slit position.  Final spectra were obtained by combining the
extracted, wavelength--calibrated pairs from the two beam positions
along the slit.  Slopes in the continuum of each source were removed
by normalization to a linear fit across the spectral interval.  Figure
\ref{fig-allspectra} displays the reduced spectra obtained in our
Ophiuchus survey.

\section{Spectroscopic Analysis}
\label{sect-spect.analysis}

We have developed a spectral fitting technique that provides
measurements of physical parameters in PMS stars (DJ03).  We use
information about the line shapes and relative strengths to derive
effective temperature, $v\sin i$ rotational broadening, and continuum
veiling from fits to spectral synthesis models and are able to measure
surface gravities from the equivalent width ratios between the Na
lines at 2.2 $\mu$m and $^{12}$CO lines at 2.3 $\mu$m.  Since we
compare our spectra to photospheric models with different effective
temperatures, we measure T$_{\rm eff}$ directly without introducing
additional uncertainty due to conversion from spectral type.  Our best
fit stellar parameters are determined by the position of the minimum
in the error space of residuals to the model fits.

\subsection{Effective Temperature, Rotation, and Veiling}
\label{sect-technique}

We outline briefly here the technique we have developed for obtaining
stellar parameters from high resolution near--IR spectra.  DJ03 discuss
details of the technique, its accuracies, and its susceptibility to
systematic errors.  We use a wavelength region in the K--band at
2.200--2.210 $\mu$m (see Fig.~\ref{fig-gss28.paramfits}) to analyze
stellar properties by examining the behavior of strong, temperature
sensitive, photospheric absorption lines (e.g. Na I, Si I, \& Sc I).
In general, as temperature decreases (T$_{\rm eff}$ = 4400--3000 K),
the wings of the Na I lines broaden and the Sc I lines grow deeper,
while the Si I lines decrease in strength.

For this given wavelength region, we develop a grid of synthetic
spectra to which to compare our data and determine stellar properties.
The synthetic spectra are generated from NextGen stellar atmosphere
models \citep{hauschildt1999} using the spectral synthesis program,
MOOG \citep{sneden1973}.  The grid of synthesis models covers a range
in cool effective temperatures (T$_{\rm eff}$ = 3000--5000 K) and PMS
surface gravities ($\log g$ = 3.5--5.0).  We assume solar metallicity
in generating the spectra, consistent with the results of
\citet{padgett1996}, who measures [Fe/H]~$=+0.08\pm 0.07$ for a sample
in $\rho$ Oph \citep[see also the results of][]{balachandran1994}.

We add rotational broadening ($v\sin i$ = 4--40 km~s$^{-1}$ in steps
of 1 km~s$^{-1}$) and 2.2 $\mu$m continuum veiling (r$_{\rm Na}$ =
0--4.0 in steps of 0.1) to these model spectra and compare each
modified spectrum to the observed spectrum at a given radial velocity.
We define the 2.2 $\mu$m veiling as the ratio of continuum excess at
2.2 $\mu$m (the wavelength of the Na interval) to the continuum
arising from the stellar photosphere (r$_{\rm Na}$ $\equiv$ F$_{2.2
\mu \rm m~excess}$/F$_{2.2 \mu \rm m~stellar}$).  A range of radial
velocities ($\pm$ 30 km s$^{-1}$) is tested by shifting the model in
increments of 0.8 km s$^{-1}$ (0.5 pixels).  We assume the infrared excess is
uniform in intensity across our wavelength interval at 2.2 $\mu$m.
The best model spectrum is selected by finding the minimum RMS of the
residuals to the fit from selected wavelength sub--intervals that are
centered on the Na I lines (see vertical dashed lines
Fig.~\ref{fig-gss28.paramfits}).

We start with an assumed gravity of $\log g$ = 3.5, which corresponds
to $\sim$1--2 Myr old objects in stellar evolutionary models
\citep{baraffe1998, siess2000, palla2000}, consistent with age
estimates of the central embedded cluster in Ophiuchus
\citep{wilking1989, greene1995, luhman1999, bontemps2001}.  Next, we
compare the observed spectrum to models over the entire range of our
synthesis grid in effective temperature, $v\sin i$ rotation, and
continuum veiling.  We then find the best fit values of these
parameters by examining the RMS deviation of the models from the
observed spectrum over the regions where lines are present (see DJ03),
and then interpolating between these models to find the values with
the minimum RMS.  Figure~\ref{fig-gss28.paramfits} illustrates our
spectral fitting technique as it applies to one of the PMS sources in
our sample (GSS 28).

\subsection{Quality of the Model Fits}

The random errors in the model fits are small.  We estimate these
errors by finding the values of the parameters for the minimum RMS
when comparing our synthesis grid to target model spectra to which we
have added Gaussian noise (DJ03).  By looking at the variation in the
values of the parameters at the point where the RMS is minimized for
many independent Gaussian noise seeds (at S/N = 30 per pixel, R$_{\rm
pix}$ = 240,000), we find ($\pm$1 $\sigma$) uncertainties of T$_{\rm
eff}$ = 45 K, r$_{\rm Na}$ = 0.04, and $v\sin i$ = 1.9 km s$^{-1}$.

At a S/N of 30, errors in our fits for temperature are dominated by
systematic effects.  When we compare effective temperatures of MK
standards derived by our near--IR spectral fitting technique to values of
T$_{\rm eff}$ derived by converting optical spectral types to
temperature, we find agreement to a level of $\pm$140 K, 1$\sigma$,
consistent with the uncertainties involved in conversion of spectral
type to effective temperature.  Tests of the fitting procedure run on
MK standards to which artificial infrared excesses had been added
indicated a small systematic overestimate of veiling $\Delta$r$_{\rm
Na}$= 0.13$\times$(1+ r$_{\rm Na}$)(DJ03).

Rotation, temperature, and gravity affect the shapes of photospheric
lines in different ways.  Fully resolved spectra of transitions with
different oscillator strengths in species with different energy levels
and different abundances can therefore separate the effects of changes
in $v\sin i$, T$_{\rm eff}$, and $\log g$, even when equivalent width
ratios give ambiguous results.  Nevertheless, there are some
combinations of these parameters for which the RMS difference between
the synthetic spectra in the grid and the target spectrum do not
change very quickly.  The most notably troublesome combination is
temperature and gravity (see $\S$ \ref{sect-spec.logg}), which we
handle by making use of additional information about the (2--0)
$^{12}$CO bandhead absorption, when available.  The quoted random
errors were derived holding $\log g$ fixed at 3.5 while allowing the
remaining three parameters to vary independently and therefore reflect
the presence of regions in the parameter space with unusually slow
variations in the RMS difference between observed and model spectra as
pairs of parameters vary simultaneously.

Historically, ultraviolet and optical continuum excesses of T Tauri
stars (TTSs) have been determined by using the spectra of
non--accreting, non--optically veiled stars as a template spectra, as
opposed to the synthetic models which we use as template spectra.  In
both cases the continuum excesses are inferred directly from the
diminished line--depths of photospheric features \citep[e.g.][]
{hartigan1989, hartigan1991, basri1990}.  These veiling studies, which
use observed template spectra, have demonstrated that the veiling
measurements are sensitive to the choice of the adopted template.  Our
method of generating synthetic spectra, however, accounts for the
lower--than--dwarf surface gravity and determines the temperature
directly from the spectrum, thereby minimizing template mismatches.
We demonstrated that our synthetic templates are able to reproduce the
observed spectra of 9 main--sequence stars and 2 luminosity class IV
stars (assumed log g = 3.5) without any artificial veiling (DJ03).
Nevertheless, one may worry that the adopted synthetic model is too
simplistic for a T Tauri photosphere, which may have star spots and be
chromospherically active.  However, neither of these effects are
expected to significantly alter our results since the effects of star
spots are minimal at near--IR wavelengths
\citep{fernandez1996,eiroa2002}, where their flux contrast is
minimized, and the atomic lines used in the analysis here are not
chromospherically sensitive, as best we know.  Unfortunately we do not
have non--veiled, non--accreting TTS spectra in hand to directly test
synthetic models in these cases.  We note however, that the optical
spectra of non-- (or very little) veiled TTSs are reproducible with
synthetic models \citep{johnskrull1999}, which suggests that any
possible unaccounted for photospheric effects are not likely to bias
our inferred properties significantly.

\subsection{Magnetic Fields}
\label{sect-mag.fields}

The role of magnetic fields in the evolution of young stars may be
significant.  Strong magnetic fields have been measured in a small
number of TTSs \citep{johnskrull1999,johnskrull2001b}, by modeling the
shape of stellar absorption lines that have been broadened by Zeeman
splitting.  A typical disk--averaged field strength of 2.5 kG is found
from magnetic field measurements in six TTSs.

Lines that are broadened due to Zeeman splitting can introduce a bias
into the derived stellar properties if the effects of magnetic fields
are ignored in the analysis.  Since our spectral synthesis models do
not have magnetic field strength as a free parameter, we investigate
how Zeeman broadening changes the derived stellar parameters using
synthesis models that do account for the effects of magnetic fields on
the shapes and depths of the lines in the Na interval.  We quantify
the amount of change in the derived effective temperature, the
gravity, the $v\sin i$ broadening, and the veiling that a strong
magnetic field can cause by using our analysis procedure to derive
physical parameters for artificial target stars with a temperature of
3600 K, similar to the temperature of most of the sources in our
sample.  The two targets were synthetic spectra provided by Chris
Johns--Krull.  One synthetic target had no magnetic field and the
second had an average field of 2.0 kG.  We derived T$_{\rm eff}$ and
$v\sin i$ for these targets in the usual way, by finding the position
of the minimum RMS difference between these artificial stars and our
model spectra.  The best fit for the target with a 2.0 kG field had a
lower temperature and a higher $v\sin i$ value than the target with no
magnetic field by 6\% ($\sim$1.5 spectral subclasses at M0) and 10\% (2 km
s$^{-1}$), respectively.  If 2.0 kG is a typical disk--averaged
magnetic field for the sources in our sample, then our derived
temperatures are underestimated and rotational velocities are
overestimated by a similar factor.

We also examined how changes in the strengths of lines in the Na
interval affect the value of $\log g$ derived spectroscopically.  We
find the best fit to a synthetic target star (see above) that has an
average magnetic field of 2.0 kG using our models with no assumed
magnetic fields and allowing only gravity to be a free parameter.  We
measure a 16\% increase in the integrated equivalent width across the
Na interval when a 2.0 kG field is introduced, which causes a decrease
in the equivalent width ratio to the magnetically insensitive
$^{12}$CO interval (see $\S$ \ref{sect-spec.logg}).  We find that an
increase in equivalent width in the Na interval due to the presence of
a 2.0 kG magnetic field could cause an overestimates in the gravity by
$\sim$0.3--0.4 dex.  Since the magnetic field strengths for the stars
in our sample are not known, we have not corrected for this possible
effect.
   
\subsection{Spectroscopic Determination of Log G}
\label{sect-spec.logg}

For late--type stars, the RMS difference in the 2.2 $\mu$m region
between data and families of models in the temperature--gravity plane
running in the direction from lower temperature and lower gravity to
higher temperature and higher gravity changes very slowly.  If we were
to adopt $\log g$ values that are larger or smaller by 0.5 dex than
the value adopted here (see $\S$ \ref{sect-technique}), the derived
temperatures will change by $\pm \sim$200 K and yield similar quality
fits see (Fig.~6 in DJ03).  To resolve this ambiguity, we can use
spectra from the (2--0) $^{12}$CO line region (2.2925--2.2960 $\mu$m).
The strength of the $^{12}$CO bandhead is a sensitive luminosity
diagnostic in late type stars \citep{baldwin1973, kleinmann1986,
ali1995, ramirez1997}.  CO is only mildly sensitive to changes in
effective temperature.

The ratio of photospheric line strengths between the Na and CO
spectral regions depends strongly on gravity at a given temperature
but is only weakly dependent on temperature at a fixed gravity.  We
can therefore employ this ratio as an indicator of surface gravity in
the two stars for which we have measurements of both spectral regions,
GY 314 and IRS 2.  We first find the best temperature at an assumed
$\log g$ of 3.5 using our spectral fitting routine. For these T$_{\rm
eff}$ values, we then use the $^{12}$CO/Na line strength ratio
(Fig.~\ref{fig-isograv}) to estimate the gravity.  We iterate between
determinations of T$_{\rm eff}$ and $\log g$ until the values converge
(DJ03).  This method gives $\log g$ values of 4.3 $\pm$0.09 for IRS 2
and 3.5 $\pm$0.10 for GY 314.  However, the model atmospheres from
which the comparison spectra are derived are binned in steps of
$\Delta \log g$=0.5.  The best $\log g$ value for IRS 2 lies in between
two models on the grid.  Since the behavior of the line shapes at 2.2
$\mu$m is nonlinear with changes in surface gravity, we adopt $\log
g$=4.0, which has the lowest RMS, rather than interpolating between
models with $\log g$=4.0 and $\log g$=4.5.  This choice likely has the
effect of lowering the determined temperature for IRS 2 by $\sim$180
K, about 1 spectral subclass.

\section{Results}
\label{sect-results}

\begin{figure}
  \epsscale{0.8} \plotone{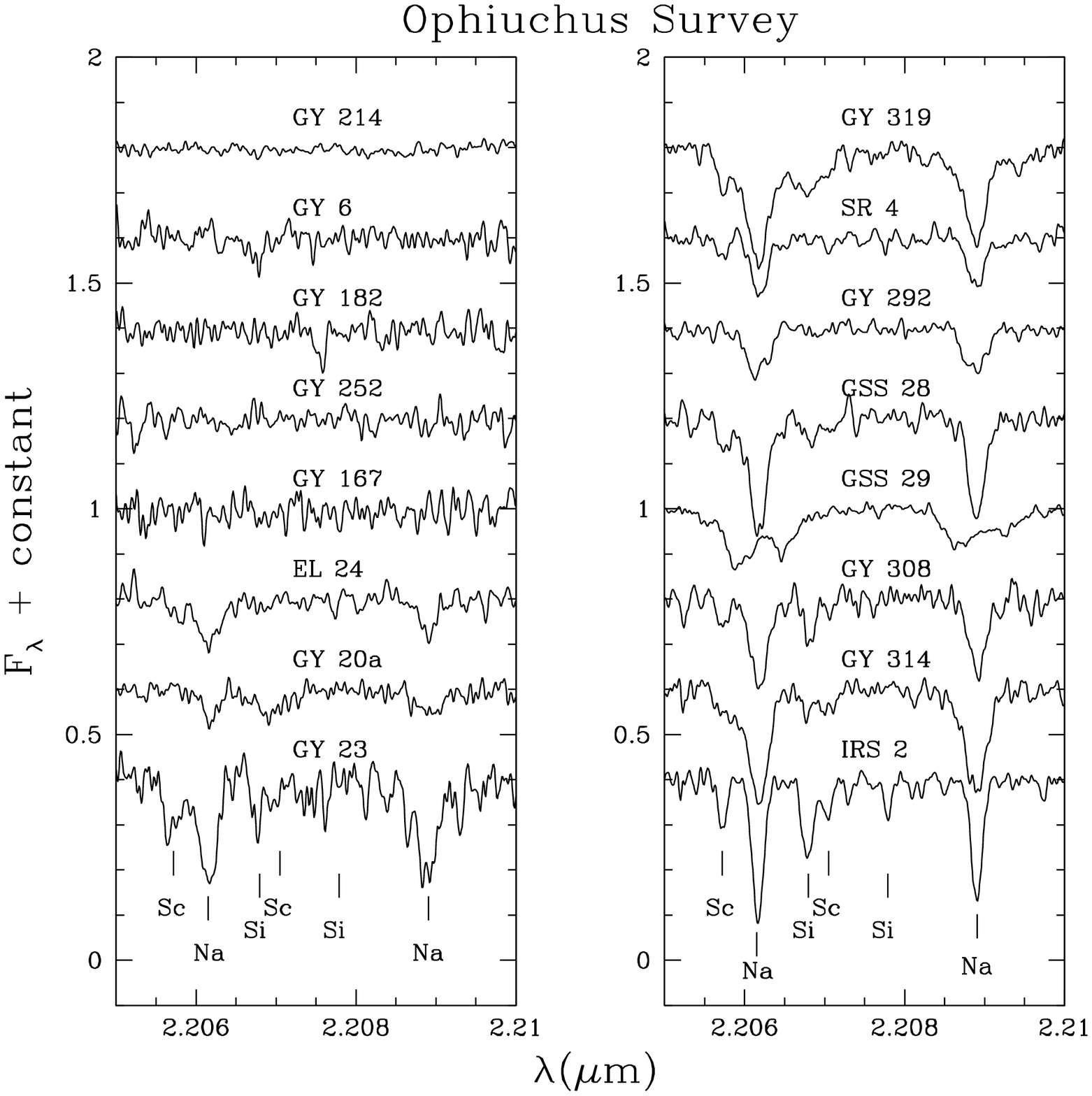}
\caption[Near--IR spectra in the K--band of our Ophiuchus sample of PMS
stars] 
{\label{fig-allspectra} K--band spectra of Ophiuchus targets
observed at high spectral resolution (R=50,000) using the PHOENIX
spectrometer on the Kitt Peak 4--meter.  The spectra have been
continuum normalized and are offset for clarity.  The first 5 spectra
(GY~214, GY~6, GY~182, GY~252, and GY~167), show no photospheric features
at a S/N of $\geq$ 30 (see $\S$ \ref{sect-results}).  
}
\end{figure}

\begin{figure}
  \epsscale{0.8}
  \plotone{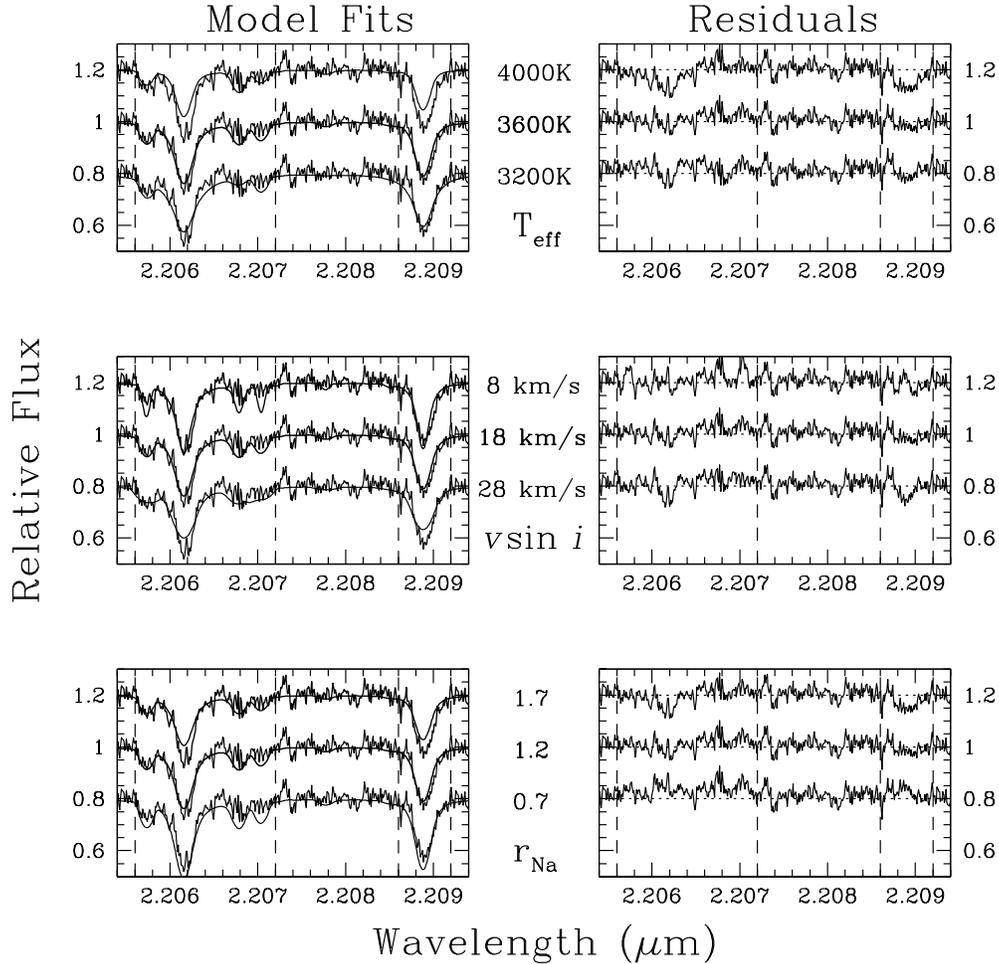}
\caption[Example of our spectral synthesis fitting technique]
{\label{fig-gss28.paramfits} Spectral synthesis fits to the observed
spectrum of GSS 28 (left panels) and residuals to each fit (right
panels).  The variation of line shapes with different effective
temperatures (top left panel), different $v\sin i$ rotations (middle
left panel), and different veilings (bottom left panel) is visible.
The best fit to each stellar parameter (holding the other parameters
fixed) is the middle spectrum in each of the three panels.  The best
fit model corresponds to $v\sin i$ rotation of 18 km~s$^{-1}$, veiling
of 1.2, and effective temperature of 3600 K.  The dashed lines in the
panels indicate the two regions containing photospheric absorption lines
where the RMS of the residuals
is computed in each plot (2.2056--2.2072 $\mu$m and 2.2086--2.2092
$\mu$m).}
\end{figure}

\begin{figure}
  \epsscale{0.8} \plotone{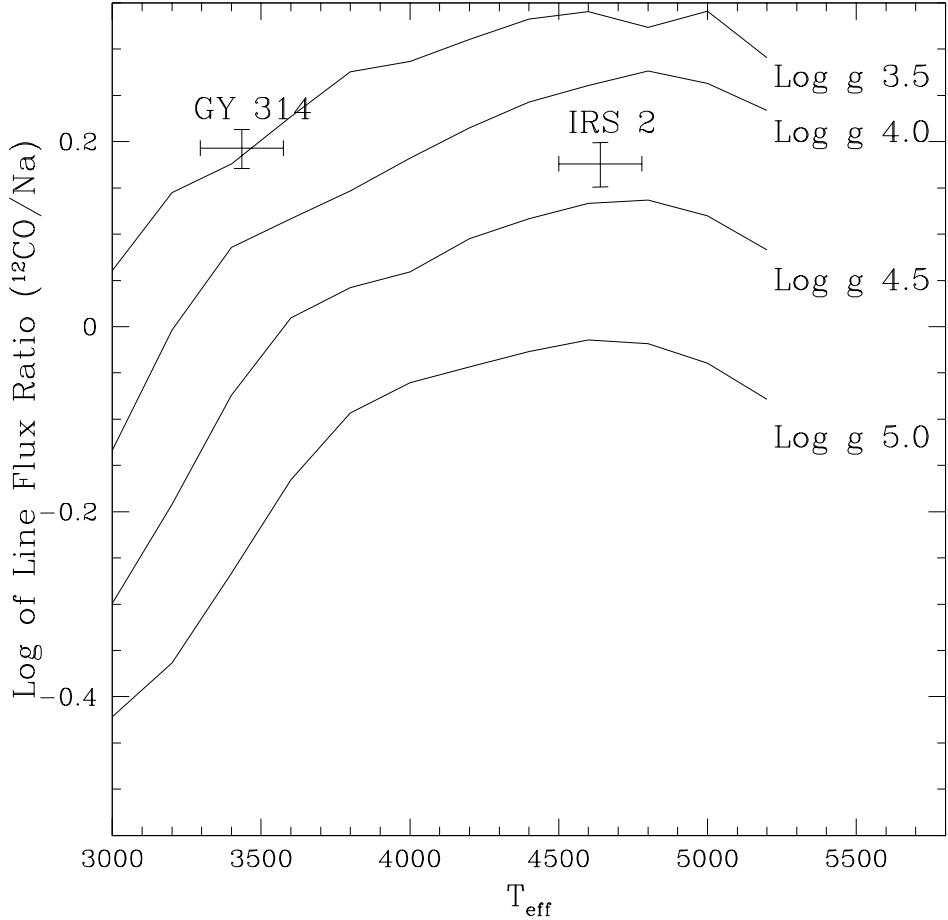}
\caption[Isogravity contours for the $^{12}$CO/Na line flux region]
{\label{fig-isograv} Isogravity contours as a function of
the ratio of $^{12}$CO/Na line flux and effective temperature.
Equivalent widths were computed from 2.2020--2.2120 $\mu$m in the Na
interval and 2.2925--2.2960 $\mu$m in the $^{12}$CO interval.  We
calculate the surface gravity of 2 Ophiuchus sources from their
effective temperatures and ratio of photospheric line fluxes measured
in two regions at high spectral resolution, scaled to the slope of the
observed SED at low resolution.  T$_{\rm
eff}$ errors (horizontal error bars) are dominated by systematics in fitting
observed spectra to the synthesis models ($\pm$ 140 K, i.e. one
spectral subclass).  Errors in the measured line flux ratio (vertical
error bars) are dominated by the determination of the continuum level,
a function of S/N.  Surface gravities derived using this technique are
independent of uncertainties in the line--of--sight extinction and
continuum veiling that are problematic in photometric measurements.}
\end{figure}

Figure~\ref{fig-allspectra} shows the reduced spectra of all 16
sources we observed in the Na interval.  The two strongest absorption
features are the neutral Na lines at 2.206 and 2.209 $\mu$m.  The
Class I and I.5 sources (GY 214, GY 6, and GY 252) do not show
photospheric features at a S/N of $\sim$70.  GY 167 (Class II) is also
featureless, though at lower S/N.  GY 182, reported to be early
spectral type (B8--A7) by LR99, does not show evidence of photospheric
lines at 2.2 $\mu$m, consistent with this early spectral type.

Ten of 16 sources had detectable photospheric lines (e.g. Na I, Si I,
\& Sc I) to which we could apply our fitting technique.  The derived
properties are listed in Table~\ref{tbl-oph.params}.  These 10 sources
have a radial velocity consistent with that of the $\rho$~Oph cloud
\citep[v$_{\rm LSR}$ of 2.8--4 km
s$^{-1}$,][]{encrenaz1974,myers1975}.  The mean radial velocity
(v$_{\rm LSR}$) of these 10 sources is 3.9 km s$^{-1}$, with a
standard deviation of 1.7 km s$^{-1}$ about the mean.

One source, GSS 29, has a radial velocity offset from the cloud of
-28.2 km~s$^{-1}$ (see Fig.~\ref{fig-allspectra}).  In this case,
fits to the photospheric lines are unsatisfactory because the
absorption lines are too broad and asymmetric, suggesting the presence
of a spectrally unresolved secondary companion.  Subsequent
observations of this source have shown it to be a radial velocity
variable.  Although we did not analyze GSS 29 with our spectroscopic
fitting routine, we conclude that it is a young spectroscopic binary
in $\rho$~Oph with a mid--K spectral type.

For eight of 10 sources with fittable lines, we measure cool effective
temperatures that are within a relatively narrow range (3400--3700 K).
The remaining two sources had temperatures roughly 1000 K hotter.
Spectral types are assigned from the derived temperatures using the
Sp--T$_{\rm eff}$ conversion of \citet{dejager1987}.  In the 10
sources with good line detections, we see a broad range in stellar
rotation ($v\sin i$ = 12--39 km s$^{-1}$) and veiling at K (r$_{\rm
Na}$ = 0.3--4.5).

For two sources (GY 314 and IRS 2), where high resolution spectra
including the (2--0) $^{12}$CO bandhead are available, we employ our
full fitting technique ($\S$ \ref{sect-spec.logg}) that uses both
$^{12}$CO (2.293 $\mu$m) and Na I (2.207 $\mu$m) line regions to
measure the surface gravity, spectroscopically.  For the remaining
eight sources, we assumed a surface gravity of $\log g$ = 3.5.
Table~\ref{tbl-oph.params} summarizes all the stellar parameters we
measured in our Ophiuchus sample.

\subsection{Luminosities Determined from Photometry}
\label{sect-lums.from.phot}

We derive photospheric luminosities for our sources listed in
Table~\ref{tbl-oph.params} in two ways, both based on the near--IR
photometry of \citet{barsony1997}.  First, we use the published
J--band apparent magnitudes and assume there is no excess present.  We
then correct for the extinction at J by using the observed J--H
colors, adopting the extinction law of \citet{martin1990}, and
applying our measured stellar temperatures to the color--temperature
relation in Table A5 of \citet{kenyon1995}.  Using a distance modulus
of 5.81 for Ophiuchus \citep{dezeeuw1999}, we obtain the absolute
magnitude at J.  We convert to bolometric luminosity using the derived
effective temperature from our best model fit to the data and the
bolometric correction factor from Table A5 of \citet{kenyon1995}.

\begin{figure}
  \epsscale{0.8} \plotone{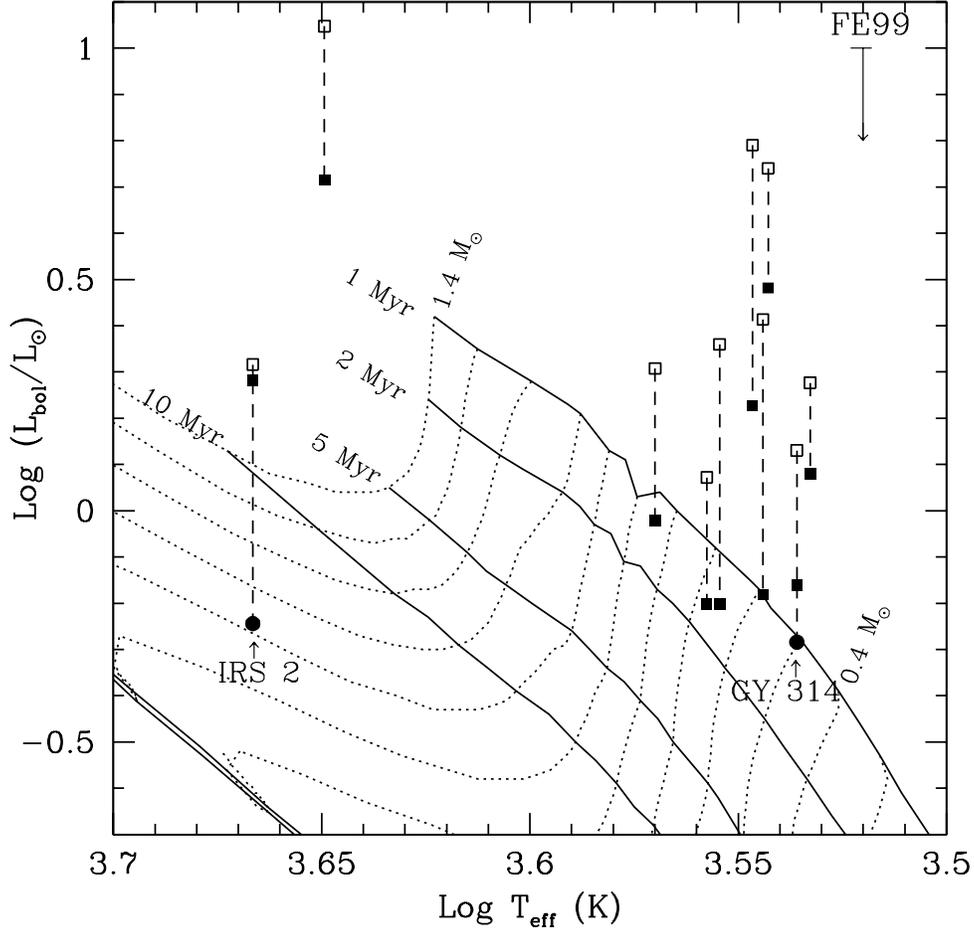}
\caption[Derived luminosities and temperatures for our PMS sources
plotted with Baraffe stellar evolutionary models]
{\label{fig-hrchange.pms} Differences in derived photometric
luminosities of Class II PMS objects in Ophiuchus 
superposed on theoretical evolutionary tracks \citep{baraffe1998}.  The open
squares show the luminosities determined from J--band photometry with an
assumed excess at J of zero.  The filled squares show the same sources
(connected by dashed lines) with luminosities calculated from K
photometry, and accounting for our measured K--band excesses.
Additionally, the solid circles show luminosities determined from
spectroscopy for two sources (labeled).  The solid vertical bar in the
upper right corresponds to a downward shift in the derived luminosity
when the J--band veiling increases from zero to r$_{\rm J}$ = 0.57, the
average veiling value at J seen by \citet{folha1999} in their sample
of TTSs.}
\end{figure}

Our second way of deriving luminosities from the photometry follows
the same procedure as above except that we use the published K--band
apparent magnitudes and remove the flux that is due to the continuum
excess in the K--band that we have measured from our best model fits
to the spectra.  The K--band flux from the stellar photosphere is
F$_{\rm K*}$=F$_{\rm Ktot}$/(1+r$_{\rm Na}$), where F$_{\rm Ktot}$ is
the flux corresponding to the apparent magnitude of the source.  We
then compare the luminosities derived in these two ways by plotting
these sources against their measured effective temperatures on an H--R
diagram together with the model tracks of \citet{baraffe1998}
(Fig.~\ref{fig-hrchange.pms}).

The dashed lines in Figure \ref{fig-hrchange.pms} connect the stellar
luminosities derived from the dereddened J--band magnitudes to the
luminosities derived from the dereddened K--band magnitudes from which
we have subtracted the non--photospheric excess using the values of
r$_{\rm Na}$ derived from our high resolution spectroscopy.  The
luminosities based on the J--band photometry are systematically higher
(by an average of 0.35 in the log) than the luminosities derived from
the corrected K--band magnitudes.

\subsection{Luminosities Determined from Spectroscopy}

For IRS 2 and GY 314, where we were able to determine $\log g$
directly from spectra ($\S$ \ref{sect-spec.logg}), we can determine
the luminosity independent of the K--band photometry if the stellar
mass is known.  We use the PMS model tracks of \citet{baraffe1998} to
fit our observations of $\log g$ and T$_{\rm eff}$ to select a unique
mass track.  Using the relationship between gravity and mass, we
derive a luminosity:

\begin{equation}
\label{eqn-lum.grav}
L = 4\pi \sigma G \frac{MT^4_{\rm eff}}{g}
\end{equation}

Since conversion from gravity to luminosity is dependent on a
particular evolutionary model, we derive masses and ages for IRS 2 and
GY 314 using 4 independent theoretical evolutionary tracks (see
Table~\ref{tbl-mass.age}).  For simplicity, only the Baraffe models
are used when displaying the luminosities of our sources for which we
have spectroscopically measured surface gravities.  These luminosities
along with our measured effective temperatures for both sources are
plotted on the H--R diagram against the Baraffe stellar evolutionary
model tracks (Fig.~\ref{fig-hrchange.pms}).

\section{Discussion}

\subsection{Cool Derived T$_{\rm eff}$ Values and Significant Veiling}

The majority of our sources (8/10) have temperatures near 3500 K,
corresponding to $\sim$M2 spectral type.  Seven of 8 of these cooler
temperature sources also have temperatures derived from lower spectral
resolution (R = 800--1200) observations (LR99).  In agreement with the
results presented here (DJW03), LR99 also see a narrow range in their
derived temperatures ($\sigma$ = 50 K).  The temperatures reported
from the lower spectral resolution study, however, are systematically
higher than the temperatures we derive for these seven sources at
higher spectral resolution ($<$T$_{\rm eff[LR99]}$~--~T$_{\rm
eff[DJW03]}$$>$~=~325 K).

A second systematic difference between our results and LR99 is that
the veiling values we derive are larger than those determined from low
resolution spectra.  The median value in the expression: $(1+$r$_{\rm
Na})_{\rm DJW03}$~/~$(1+$r$_{\rm K})_{\rm LR99}$ is 1.5 with a
range of 1.1 to 2.6.  This leads to a mean difference in the 2.2
$\mu$m continuum veiling of 1.2 from the two independent veiling
determinations for stars in our sample.  Part of the difference
between the low and high resolution determinations of the 2.2 $\mu$m
veiling may result from a small systematic error in the high
resolution r$_{\rm Na}$ fits.  When we artificially add continuum
veiling to the spectrum of an MK standard and then recover this
veiling using our fitting routine, the best--fit result is
systematically larger by $\Delta$r$_{\rm Na}$=$0.13 \times(1+$r$_{\rm
Na})$ than the veiling we added to the MK standard (DJ03).  If a
systematic effect of the same size is present in the analysis of Class
II sources, our analysis of the high resolution spectra overestimates
the value of r$_{\rm Na}$ by 0.15--0.7 as r$_{\rm Na}$ ranges from 0.3
to 4.5.  This effect, however, only accounts for 1/4 of the difference
between the veiling values determined from our high resolution spectra
and the low resolution spectra of LR99.  Furthermore, tests with
higher surface gravity models (log~g $>$ 4.5) show less of an offset
between our recovered veiling values and the continuum excess we added
to our late--type dwarf standards.

The largest part of the T$_{\rm eff}$ and 2.2 $\mu$m veiling
difference between our results and those of LR99 most likely stems
from the differences in spectral resolution.  At lower spectral
resolution (R $\lesssim$ 5,000), where K--band absorption lines from
stars with cool photospheres are unresolved, measurements of effective
temperatures and veiling must rely solely on equivalent widths.  As a
consequence, lower resolution measurements cannot distinguish between
temperature and veiling changes in temperature--sensitive lines
because both types of change can have similar effects on the
equivalent width.  The degeneracy between temperature and veiling is
split when lines are completely resolved and information becomes
available from the intrinsic {\it shapes} of spectral features.  We
determine more accurate veiling and temperature values from the high
resolution spectra because we can make use of both the depths of the
line cores and the shapes of the line wings.

\subsection{Luminosity}
\subsubsection{Luminosity and Age Estimates}

The luminosities derived from J magnitudes and from K magnitudes
corrected for excess emission are plotted against the effective
temperatures derived from our high resolution spectroscopy in
Figure~\ref{fig-hrchange.pms}.  The figure also shows the luminosities
determined for IRS 2 and GY 314 using our $^{12}$CO/Na line flux ratio
technique (see $\S$ \ref{sect-spec.logg}). The spectroscopic technique
derives $\log g$ rather than L.  The spectroscopic luminosity
therefore depends on the T$_{\rm eff}$--mass relation from the PMS
evolutionary models of \citet{baraffe1998} (also shown in
Figure~\ref{fig-hrchange.pms}). The conversion from $\log~g$ to mass
for other PMS models produces comparable luminosities for IRS 2 but
results in even lower values for GY 314 (Table \ref{tbl-mass.age}).
The Ophiuchus stars exhibit a large spread in luminosity, with 3500 K
stars ranging from 0.6 L$_\odot$ to 5.2 L$_\odot$ (based on the K
magnitudes).  The Class II sources in our survey correspond to
classical TTSs.  Since these stars are on the convective and primarily
vertical portion of their evolutionary tracks \citep{tout1999}, this
spread in luminosity for stars with similar values of T$_{\rm eff}$
may therefore be caused by a spread in age.  The photometric results
indicate that all the stars except IRS 2 lie above the 2 Myr
isochrone.  We note that the young age is consistent with values
determined for Ophiuchus stars studied spectroscopically and matched
to different PMS models \citep[e.g.][LR99]{greene1995}, but caution
that interpretation of properties at young ages (i.e. $<$ 1 Myr) is
subject to considerable uncertainty \citep{baraffe2002}.

Spectroscopically determined luminosities are largely unaffected by
the problems that can bias the photometric estimates (e.g. extinction,
veiling, unresolved binaries; see below).  For both IRS 2 and GY 314,
the spectroscopically determined luminosities are lower (by a factor
of 3.2 and 1.4, respectively) than the luminosities determined from
the corrected K magnitudes.  The spectroscopic ages for the two
objects also differ sharply (20 Myr vs. 1 Myr) lending credence to the
possibility that a range in ages causes the luminosity spread among
the 3500 K stars.  Although it is difficult to draw conclusions based
on only two stars, the differences between the photospheric and
spectroscopic luminosities and ages lead us to make a critical
examination of the validity of the luminosities derived from
photometry.

\subsubsection{Accuracy of Photometrically Derived Luminosities}

Analyses of the J and K--band photometry both result in higher
luminosities than the spectroscopic analysis and imply an order of
magnitude spread in luminosity for sources at 3500 K.  While a true
spread in source ages may explain the range of luminosities, we still
need to examine other possible causes for the luminosity spread and
also need to look for a reason for the offset between the
spectroscopic and photometric luminosities and the significant offset
between the J--band luminosities and the luminosities determined from
the veiling--corrected K magnitudes.

1. {\it Near--IR Photometric Variability}:~ 
Temporal variations in the near--IR energy output of a YSO, possibly due to
accretion or star spots, will increase the luminosity dispersion for a
young star sample.  However, this effect is usually not large at near--IR
wavelengths. \citet{barsony1997} measure J--band photometric
variations of $\Delta m_{\rm J}$ $>$ 0.4 (2$\sigma$) in only 6 of the
79 sources in their Ophiuchus sample for which they have multiple
measurements.  \citet{kenyon1995} find a mean variability of $\sim$0.1
magnitudes at K in a sample of PMS stars in Taurus--Auriga.

2. {\it Unresolved Binaries}:~
Binaries are common in regions of star formation, and probably even
more common among young stars than field dwarfs \citep{duquennoy1991,
ghez1993, ghez1997, mathieu2000}.  Mistaking unresolved binaries for
single stars would systematically bias the photometry to higher
derived luminosities (factor of $\leq$ 2), introducing additional
scatter.

3. {\it Continuum Excess in the J and H--bands}:~
J--band photometry has been commonly used to derive stellar luminosity
under the assumption that the accretion luminosity is negligible at
1.2 $\mu$m \citep[][LR99]{kenyon1995}.  The assumption that one could
neglect accretion luminosity was tested by \citet{folha1999} who used
1.2 $\mu$m spectroscopy to derive J--band continuum veiling in a
sample of classical TTSs in Taurus. The average derived r$_{\rm J}$
value is 0.57, with a maximum of 1.2.  The arrow in the upper right
corner of Figure \ref{fig-hrchange.pms} shows that the luminosities
derived for our sources in Ophiuchus from J magnitudes could agree
quite closely with the luminosities derived from the
veiling--corrected K magnitudes if the average J--band veiling derived
by \citet{folha1999} for their Taurus sample were removed from the
Ophiuchus sources.  Unaccounted for non--photospheric emission is a
plausible explanation for the high luminosities inferred from the J
band photometry.

If veiling is present at H as it is at J and K, and if the veiling
flux has an intrinsically red color, observers may be overestimating
the extinction and therefore overestimating the intrinsic flux when
they use the observed J--H color to deredden their sources.
\citet{meyer1997} examined a population of optically visible classical
TTSs. Assuming that the J--band excess is 10\% of the V band excess
derived spectroscopically for these stars, they calculated a mean
dereddened H--band veiling for their sample of $<$r$_{\rm
H}$$>$=0.2. \citet{folha1999} use their spectroscopic determinations
to derive a mean J--K color of 0.7 magnitudes for the veiling flux in
their sample.  The errors caused by overcorrecting for reddening based
on the J--H colors would affect our corrected K magnitudes as well as
the luminosities derived from the J magnitudes.

\subsection{Class I and I.5 Sources}

We did not detect photospheric lines in any of the Class I or flat
spectrum sources we observed.  The absence of detectable photospheric
lines could be due to rapid rotation, heavy veiling, diminished line
strengths due to higher photospheric temperature or some combination
of these factors.  It is plausible that younger YSOs still undergoing
gravitational collapse will be rotating rapidly and have significant
accretion.  \citet{greene2000} conclude that the lack of spectral
features in their sample of Ophiuchus Class I YSOs is due to heavy
continuum veiling (r$_{\rm K} \geq 4.0$) and rapid rotation.  The
broadened lines detected in some of their flat spectrum sources is
consistent with rapid rotation and is in agreement with the $v\sin i =
50$ km s$^{-1}$ measured in one Class I source (YLW 15) which has a
veiling of r$_{\rm K} = 3.0$ \citep{greene2002}.

One out of the five sources in our sample that had no photospheric
lines detected (GY 182) was reported to be of early spectral type
(B8--A7), while the remaining others had no reported spectral type by
LR99.  It is possible that the other sources have no observable
K--band lines solely because their photospheres are hot (T$_{\rm eff}
> 5000$ K).  In this case, no veiling or rotation would be required to
produce a featureless K--band spectrum in hot stars.

\begin{figure}
  \epsscale{0.8}
  \plotone{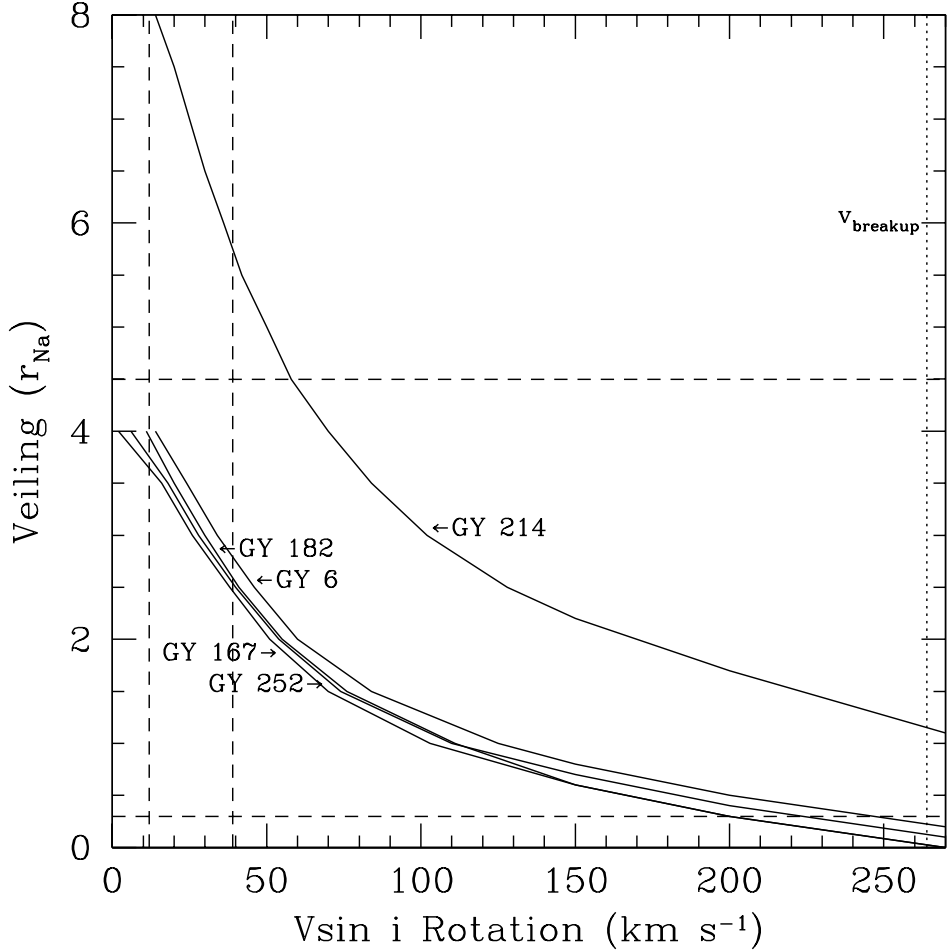}
\caption[$V\sin i$ lower limits for spectra that had no absorption
features detected] 
{\label{fig-limits.nolines} Detection limits for the 2.2 $\mu$m
Na lines as a function of stellar rotation ($v\sin i$) and continuum
veiling (r$_{\rm K}$) for the 5 PMS sources that appeared featureless
at 2.2 $\mu$m.  Each source would have had detectable Na
lines if T$_{\rm eff}$=3600 K and the veiling and rotation placed it
to the lower left of the curve.  The minimum rotation and veiling
solutions (solid lines) are set by the RMS of the residuals in the Na
line cores to the model fits being 3 $\sigma$ above the noise in the
continuum.  A hard upper limit to $v\sin i$ is set by the breakup
velocity (dotted line).  The full range we derive for rotation
(vertical dashed lines) and veiling (horizontal dashed lines) of the
Class II sources (Table~\ref{tbl-oph.params}) with measureable lines
in our sample are shown for comparison.}
\end{figure}

To investigate whether rotation and veiling could explain the lack of
detectable lines in our sample of Class I and flat spectrum sources,
we set detection limits by finding the maximum veiling as a function
of $v\sin i$ rotation allowed for at least a 3 $\sigma$ detection of
the cores of the Na lines (Fig.~\ref{fig-limits.nolines}), assuming a
temperature of 3600 K.  If these stars rotate $\leq$ 50 km~s$^{-1}$,
then significant veiling (r$_{\rm K}$ $\geq$ 2.0 for all sources and
r$_{\rm K}$ $\geq$ 4.5 for GY 214) is needed in order to hide their
photospheric lines below a 3 $\sigma$ detection limit.

\subsection{$^{12}$CO Line Contributions from the Disk }

We rely on $^{12}$CO equivalent width measurements to determine the
surface gravity in objects for which we have high resolution
observations in two wavelength intervals (2.2 and 2.3 $\mu$m, see $\S$
\ref{sect-spec.logg}).  While some models predict the presence of
$^{12}$CO absorption in the disks of TTSs \citep{calvet1991}, the most
definite detections of the CO overtone bandheads from disks around
TTSs have been in emission \citep[][LR99]{thompson1985, carr1993,
najita1996}.  This $^{12}$CO bandhead emission is rare and only seems
to be associated with higher luminosity objects.  We must avoid using
our spectroscopic method on stars that exhibit CO emission.  The
presence of even weak emission may be easy to notice in some cases,
given that high spectral resolution measurements of WL16 show CO
emission lines broadened by up to 400 km s$^{-1}$ \citep{dent1991}.
However, \citet{carr1992}, see CO emission from an outflow in the
infrared source, SVS 13 with lines that are only 40 km s$^{-1}$ wide.
$V\sin i$ broadening in CO absorption lines of some classical TTSs is
consistent with stellar rotation rather than Keplerian motion in a
circumstellar disk \citep{casali1996}.  If the origins of CO emission
and absorption at 2.2935 $\mu$m are from different velocity fields
(Keplerian disk vs.  stellar photosphere), then at resolving powers of
R $>$ 10,000, the presence of both components would be detectable in
the spectrum.  \citet{johnskrull2001a} find good agreement between
their $v\sin i$ measurements of resolved $^{12}$CO R branch lines in
the K--band to the rotational broadening measured in the optical.  For
the two measured sources in our sample, the $^{12}$CO and Na line
widths are also consistent. We therefore believe that the $^{12}$CO
absorption we are using to determine $\log g$ accurately reflects the
line flux arising in the photosphere.

\subsection{Accretion}

Class II YSOs exhibit significant amounts of near--IR emission in excess of
the expected emission from the stellar photosphere.  This excess is
apparent in the near--IR colors, where the colors of dereddened Class II
sources form a sequence in the J--H, H--K color plane that is distinct
from the colors of main sequence and post--main sequence stars
\citep{meyer1997}. Measurements of the depths of photospheric
absorption features go farther in showing that this near--IR excess is not
accompanied by significant line absorption \citep{greene1996,
casali1996, johnskrull2001a}.  \citet{greene1996} see the near--IR excess
in Class II sources as part of the overall evolution of the spectral
energy distribution.  They find no detectable lines in the K--band at
R~=~500 in Class I sources, implying that the rising near to mid--IR
spectral energy distributions (SEDs) of these objects are accompanied
by very high near--IR excesses.  The Class II objects have significant
excesses in the K--band while Class III sources have no detectable
excess.

The increasing ratio of excess to photospheric emission with
wavelength makes it clear that the near--IR excess, at least at wavelengths
beyond 1.2 $\mu$m, is not simply a long--wavelength extension of the
UV excess seen in classical TTSs.  The lack of significant line
absorption from the source of the excess makes it impossible to invoke
continuum opacity in the gas present in the hotter parts of any
accretion disk as the source of the excess \citep{greene1996,
calvet1997, johnskrull2001a}.  Our own data agree with this in showing
no absorption lines in the Class I sources and significantly weakened
but narrow absorption lines in the Class II sources.

The character of the near--IR excess emission makes it likely that it comes
from warm dust.  The problem with this idea is that it requires both a
substantial illuminating luminosity or viscous dissipation to provide
the energy and that some of the dust must be extremely hot to produce
the shortest wavelength part of this excess.  \citet{johnskrull2001a}
have examined these issues in detail and conclude that, in their
sample of classical TTSs, the measured accretion rates do not result
in sufficient energy from either source for any reasonable kind of
disk to produce the observed near--IR excess.  They argue that the strong
near--IR excesses require the presence of a significant amount of dust
surviving in the magnetosphere.

\begin{figure}
  \epsscale{0.8}
  \plotone{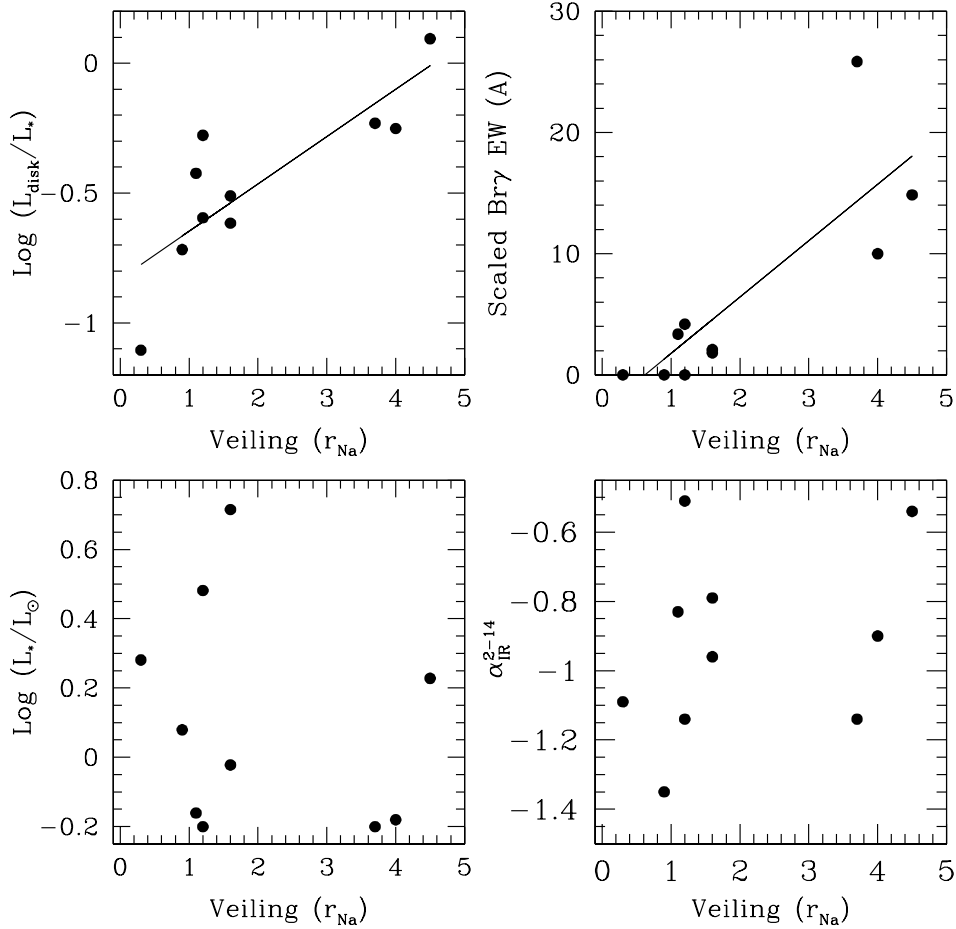}
\caption[Derived continuum veiling vs. disk properties]
{\label{fig-veiling.trends} 
The K--band veilings derived from our spectral survey in Ophiuchus vs.
(a) the ratio of mid--IR disk luminosity \citep{bontemps2001} to
stellar luminosity (upper left), (b) Brackett $\gamma$ equivalent
width (LR99) scaled by our derived K--band veiling (upper right), (c)
stellar luminosity derived from K--band photometry corrected for
continuum excess (lower left), and (d) the spectral energy index
($\alpha \equiv dlog(\lambda~F_{\lambda}) / d\lambda$) derived from 2 and
14 $\mu$m photometry \citep{barsony1997, bontemps2001} (lower right).
The solid lines in the top panels show a linear least squares best fit
to the data point.  In all panels, the x position of the points shows
the 2.2 $\mu$m veiling (r$_{\rm Na}$) determined from our spectral
fitting technique.}
\end{figure}

For our sample of Class II sources in $\rho$~Oph, we have measurements
of the characteristics of the underlying star, as well as a
determination of the amount of 2.2 $\mu$m excess from our own
observations. Measurements of Br$\gamma$ flux and mid--IR continuum
are also available for these sources \citep[LR99,][]{bontemps2001}. In
Figure \ref{fig-veiling.trends}, we compare the 2.2 $\mu$m veiling
(r$_{\rm Na}$), to various other observed parameters of the Ophiuchus
Class II sources.

The most striking result in Figure \ref{fig-veiling.trends} is the
strong correlation between strength of the near--IR veiling (r$_{\rm Na}$),
and the ratio of disk luminosity calculated from the 7 $\mu$m and 15
$\mu$m ISOCAM measurements, corrected for the contribution of the
stellar photosphere, to the stellar luminosity determined from the
veiling--corrected K magnitude. The upper left panel of Figure
\ref{fig-veiling.trends} displays this relation.  The observed
correlation shown in the upper left panel (correlation coefficient =
0.82) argues that both the near and mid--IR excess depend primarily on
the accretion rate rather than on variations in the structure of the
disk.

Since most of our sources are faint or invisible at optical
wavelengths, it is not possible to use the UV excess to directly
measure the strength of the emission produced by the accretion shock.
\citet{muzerolle1998b} have shown, however, that the Br$\gamma$ line
emission from classical TTSs can serve as a proxy in determinations of
the accretion luminosity.  The upper right panel of Figure
\ref{fig-veiling.trends} plots a quantity proportional to the ratio of
Br$\gamma$ flux to 2.2 $\mu$m photospheric flux (the product of the
measured Br$\gamma$ equivalent width and 1+r$_{\rm Na}$) against
r$_{\rm Na}$, the ratio of non--photospheric to photospheric flux at
this wavelength.  The figure shows a good correlation between
EW(Br$\gamma$)$\times$(1+r$_{\rm Na}$) and r$_{\rm Na}$ but with a
negative intercept.  One possible interpretation of this result is
that both the photosphere and the accretion shock heat the dust in the
inner disk and magnetosphere.  At low accretion rates, there is still
some excess produced by the absorption of the photospheric emission.
As the accretion luminosity grows, the amount of reprocessed radiation
emerging in the near--IR increases while the flux of the photosphere
remains constant.

The two remaining panels in Figure \ref{fig-veiling.trends} show
interesting negative results.  The lower left panel shows that the
veiling is independent of the stellar luminosity.  Since 8/10 of our
sources have a very similar T$_{\rm eff}$, this result implies that,
at least within the Class II sources, there is no clear trend relating
the veiling to the apparent age.  A more reasonable interpretation
might be that the variation in veiling in a sample with similar ages
can swamp any age--dependent secular trend in the amount of excess
emission.

The lower right panel tells a similar story.  While there is a clear
trend of increasing near--IR veiling going from Class III to Class I
sources, within Class II, there is no significant correlation between
the veiling and the 2--14 $\mu$m spectral index, the parameter most
commonly used to distinguish between the SED classes.

\subsection{Rotation}

The modest rotational velocities of classical TTSs \citep[$<v\sin i>~
\sim$15 km s$^{-1}$,][]{bouvier1986}, despite the strong evidence that
these stars were accreting material from circumstellar disks, was
initially something of a puzzle.  Models in which a strong magnetic
field couples the star to the inner edge of a truncated accretion disk
were suggested about a decade ago in order to address this problem
\citep{konigl1991,shu1994}. In this picture, kilogauss scale magnetic
fields can lock the stellar rotation rate to the Keplerian rotation of
the disk at a distance of a few stellar radii where the rotation
velocity is modest.  Near--IR measurements of Zeeman broadening show that
some TTSs have magnetic fields of sufficient strength for this to work
\citep{johnskrull2001b}.

One prediction of this magnetic braking picture is that younger stars
with high accretion rates might cause the disk truncation radius to be
smaller than for TTSs and therefore to result in a higher rotation
rate than for older systems where the magnetic field truncates the
disk at larger radii. \citet{hartmann2002} concludes, however, that
the timescales for angular momentum regulation may be comparable to or
longer than the evolutionary timescales.  If the timescales are indeed
comparable, there may be no clear trend in rotation rate with age for
the youngest objects.

\begin{figure}
 \epsscale{0.8} \plotone{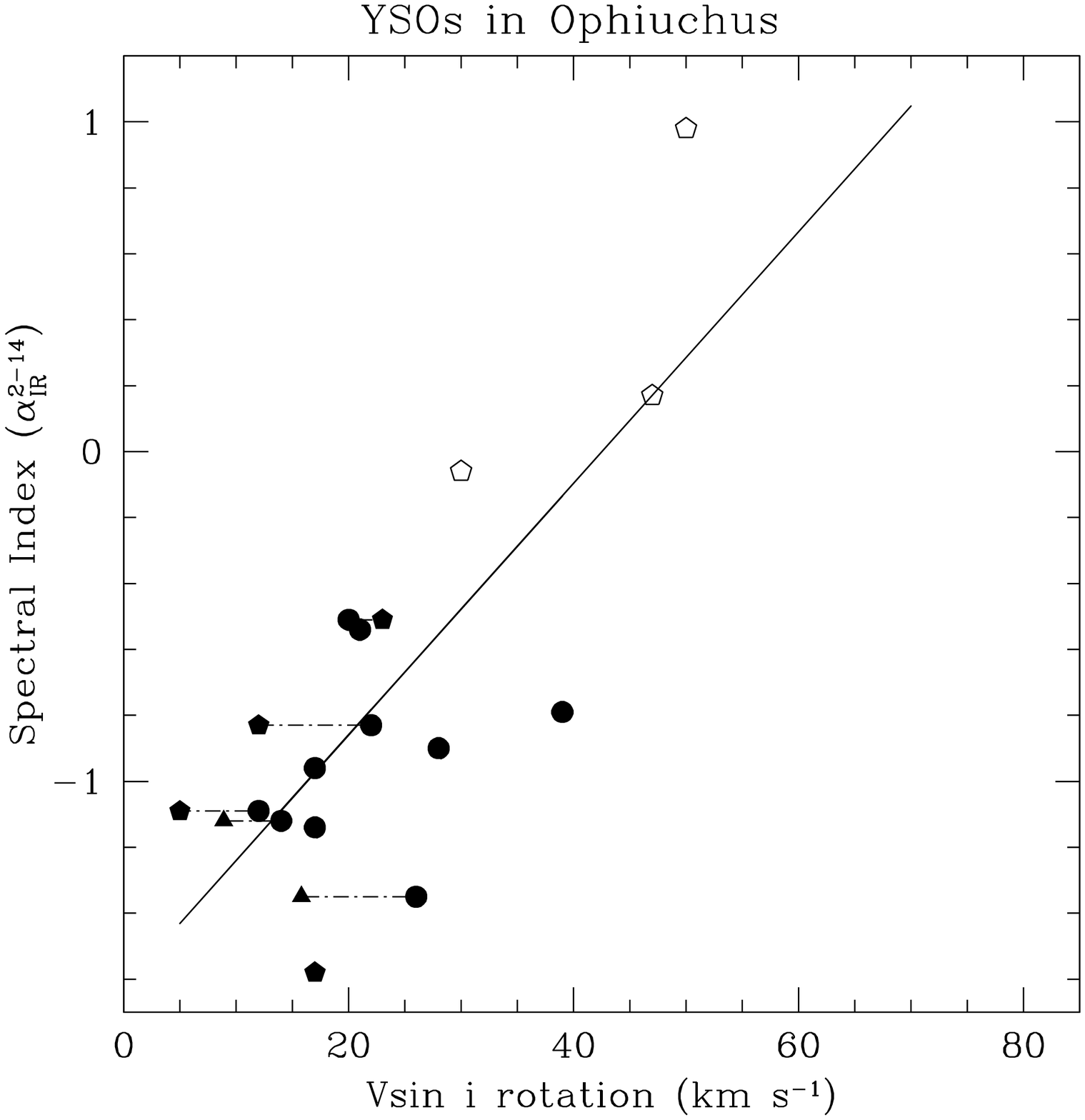}
\caption[Derived stellar rotation vs. spectral index]
{\label{fig-vsini.veil} 
Ophiuchus YSO comparisons of $v\sin i$ rotation values vs. spectral
index ($\alpha \equiv dlog(\lambda~F_{\lambda}) / d\lambda$) derived from 2
and 14 $\mu$m photometry \citep{barsony1997, bontemps2001}.  Class I
and flat spectrum sources are open symbols.  All Class II sources are
filled symbols.  The Class II sources from this study are filled
circles.  Sources from \citet{greene1997} and \citet{greene2002} are
pentagons.  The triangles show the measured $v\sin i$ rotation for T
Tauri (Class II) sources from optical spectroscopy
\citep{bouvier1986}.  YSO sources with reported $v\sin i$ measurements
from more than one study have points that are connected by horizontal
dot--dashed lines.}
\end{figure}

\citet{greene1997, greene2002} have argued on the basis of the larger
$v\sin i$ values of a few flat spectrum (Class I.5) sources, a single
Class I source, and the more modest $v\sin i$ measurements of a few
Class IIs in Ophiuchus, that there in fact is an observable trend
toward slower rotation as one goes from Class I to Class II.  Our
additional $v\sin i$ data for Ophiuchus Class II sources (Figure
\ref{fig-vsini.veil}) strengthen this result.  The figure plots the
near--to--mid IR spectral index, the continuous variable used to
determine SED class, against the $v\sin i$ values for our sample,
together with the Greene \& Lada sources.  It shows a clear trend
toward higher values of $v\sin i$ for sources with more positive
spectral indices.  Among our own sample of Class II sources, which
have spectral indices between -1.6 and -1.4, however, there is little
or no correlation with $v\sin i$.

Observers using photometric periods to determine rotation rates have
also found evidence for a relationship between infrared excess (taken
in this case as an indicator of accretion rate and youth) and
rotation.  Unlike the IR spectroscopic studies, these optical
observations examine the rotation behavior for sources with and
without near--IR excesses, i.e., those on the boundary between Class II and
Class III.  \citet{edwards1993} found that most sources with periods
$\sim$ 8 days have K--band excesses while the majority of sources with
4 day periods do not. \citet{choi1996} find that a larger sample of
young stars in Orion has a bimodal period distribution and argues that
the slower peak corresponds to stars with rotation periods
magnetically locked to their inner disks.  The observed bimodal period
distribution was later called into question by observations of a
larger sample \citep{stassun1999}, although it now appears that the
higher mass stars (M $>$ 0.5 M$_{\odot}$) do have a bimodal period
distribution \citep{herbst2002}.  Among the higher mass subgroup, the
mean I--K excess of the stars with long rotation periods is
significantly larger than the excess for stars with short rotation
periods \citep{herbst2002}.

\begin{figure}
  \epsscale{0.8}
  \plotone{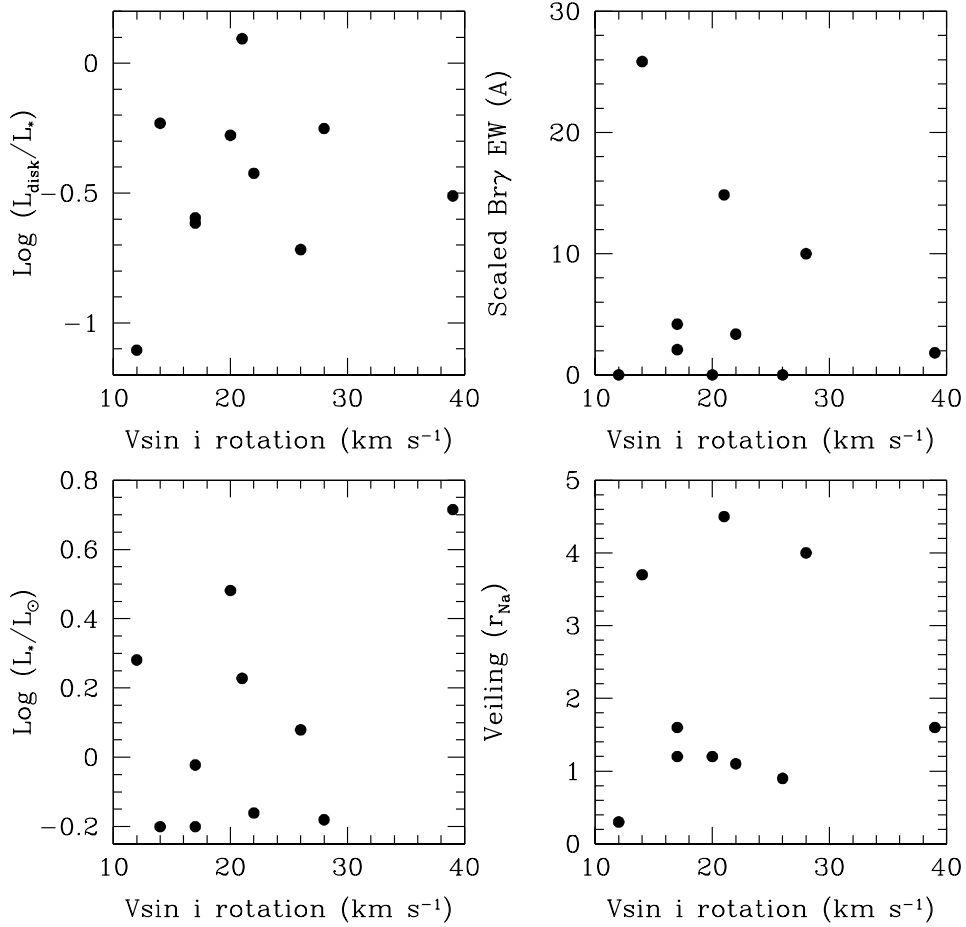}
\caption[Derived stellar rotation vs. disk properties]
{\label{fig-vsini.trends} 
Derived $v\sin i$ rotation in Class II sources vs. (a) the ratio of
mid--IR disk luminosity \citep{bontemps2001} to bolometric luminosity
(upper left), (b) Brackett $\gamma$ equivalent width (LR99) scaled by
our derived K band veiling (upper right), (c) stellar luminosity
derived from K--band photometry corrected for continuum excess (lower
left), and (d) derived continuum excess (veiling) at 2.2 $\mu$m from
our spectral fitting routine (lower right).}
\end{figure}

Previous spectroscopic studies indicate that the Class II sources have
spun down compared to the rotation of their presumably younger
counterparts (Class I and flat spectrum sources).  On the other hand,
rotation rates based on photometric periods suggest that stars without
measurable excesses at K (Class III sources) have spun up compared to
ones with significant K excesses (likely to be Class IIs).  In this
context, it is interesting to look at how various indicators of disk
accretion vary with $v\sin i$ for the Class II sources in our sample
(Figure \ref{fig-vsini.trends}).  The upper left panel shows the
variation of the ratio of disk luminosity \citep{bontemps2001} to
photospheric luminosity (this paper) which should reflect the amount
of heating of the outer disk by photospheric and accretion
luminosity. The upper right panel shows the variation of the
photospheric equivalent width of the Br$\gamma$ line which also is
related to the accretion rate \citep{muzerolle1998b}.  The lower right
panel shows the 2.2 $\mu$m veiling as a function of $v\sin i$.  There
is no clear trend in any of these panels.

The absence of a clear relation between $v\sin i$ and the various
accretion indicators in Figure \ref{fig-vsini.trends} can result from
either large variations in rotation velocity caused by inclination
effects combined with random variations in the disk emission
characteristics or from a turnover from decreasing to increasing
rotation rates occurring within the Class II population, from the
comparability of braking timescales and evolutionary timescales, or
from the contribution of factors other than magnetic locking to the
trends in rotation.  High resolution near--IR spectroscopy cannot produce
samples as large as the photometric techniques, but it can give us
rotation rates for sources across the broadest possible range in SED
classes.  An expansion of our rotation rate sample might still help to
clarify the situation regarding the evolution of rotation rates.

\section{Conclusions}

We have carried out a high resolution near--IR spectroscopic survey of YSOs
in the $\rho$~Ophiuchi dark cloud.  We measure effective temperatures,
continuum veiling, and $v\sin i$ rotation in ten Class II (T Tauri)
sources.  For two YSOs where we have high resolution spectra at 2.2
$\mu$m and 2.3 $\mu$m, we measure surface gravities using a new
spectroscopic technique we have developed and compare our results to
luminosities derived from photometry.  In five sources where we do not
detect photospheric lines, we place limits on rotation rates and the
2.2 $\mu$m veiling.

From our measurements, we draw the following conclusions.

1. With the exception of one source, GSS 29, which has a complex
spectrum, all the Class II sources in our sample have radial
velocities within a few km s$^{-1}$ of the cloud velocity determined
from millimeter molecular line measurements.  This result points out
the usefulness of high resolution spectroscopy in the near--IR for
confirming membership in the star--forming cluster.

2. Most of the brightest Class II sources in the core of the
$\rho$~Oph cloud have cool derived effective temperatures that fall
within a narrow range ($<{\rm T}_{\rm eff}>$ = 3530 K $\pm$ 100).  The
typical Class II source in our sample has a continuum excess larger
than the stellar flux at 2.2 $\mu$m ($<{\rm r}_{\rm K}>$ = 2.3).  With
high resolution spectroscopy we are able to split the degeneracy
between veiling and temperature which otherwise causes a systematic
offset to higher temperatures and less veiling, evident when we
compare our results to low resolution fitting techniques.

3. Using high resolution spectroscopy of spectral intervals including
the Na lines at 2.2 $\mu$m and the (2--0) $^{12}$CO bandhead at 2.3
$\mu$m, we derive luminosities for IRS 2 and GY 314 that are smaller
than their photometrically derived luminosities.  The ages inferred
from the spectroscopic measurements are therefore older than the ages
derived from photometry. Furthermore, these sources do not appear
coeval based on the luminosities derived by either photometry or
spectroscopy.  Note, however, that the luminosities may be
underestimated if the stars have strong magnetic fields.  Future
derivations of stellar parameters from high resolution spetra will
need to take the effects of magnetic fields into account.

4. The $v\sin i$ rotations of YSOs in $\rho$~Ophiuchi Class II sources
(this study), along with Class I.5 and I sources (other high
resolution studies) indicate that Class II YSOs have already spun down
from earlier evolutionary stages (Class I.5 \& I).

5. We do not observe any photospheric lines in the Class I or I.5
sources (4 total) we observed at S/N $\geq$ 30, either because they
have heavy veiling and/or are rotating rapidly, or have hot (T$_{\rm
  eff}$ $>$ 8000 K) photospheres.

\section{Acknowledgments}

We thank Chris Sneden for his advice and support in adapting MOOG for
use with YSOs.  We are grateful to Tom Greene for providing us with
CSHELL data on YSOs and insightful comments on this manuscript.  We
thank Ken Hinkle for help with our PHOENIX observations.  Discussions
with Isabelle Baraffe and Carlos Allende--Prieto were helpful in
preparing this manuscript.  We thank Chris Johns--Krull for insightful
comments and for generously providing us with some spectral synthesis
models that included the effects of magnetic fields.  We thank the
Max-Planck-Institut fuer extraterrestrische Physik for its hospitality
and support while parts of this project were carried out.

Early phases of this work were supported in part by NSF grant
AST95-30695 to the University of Texas at Austin.

\clearpage

\clearpage

\clearpage

\clearpage

\clearpage

\clearpage

\clearpage

\clearpage

\clearpage

\begin{deluxetable}{ccccccc}
\footnotesize

\tablecaption{Ophiuchus Survey Sample \label{tbl-oph.sample}}
\tablewidth{5.5in} 
\tablehead{
&\colhead{SED} & & & &\colhead{Obs.} &\\ \colhead{Source\tablenotemark{a}} & \colhead{Class}\tablenotemark{b} &\colhead{m$_{\rm J}$\tablenotemark{c}} & \colhead{m$_{\rm H}$\tablenotemark{c}} &\colhead{m$_{\rm K}$\tablenotemark{c}} & \colhead{Date (UT)}&  \colhead{approx. S/N}}
\startdata

GY 20A   &II     &8.9    &7.4    &6.4    &5/20/00                        &35\\
EL 24    &II     &10.2   &8.2    &6.8    &5/20/00                        &45\\
GY 167   &II     &9.8    &8.3    &7.1    &5/22/00                        &25\\
GY 23    &II     &10.8   &8.7    &7.2    &5/22/00                        &25\\
GY 319   &II     &8.5    &7.7    &7.2    &5/20/00                        &50\\
SR 4     &II     &9.0    &8.0    &7.3    &5/21/00                        &55\\
GY 168\tablenotemark{d}  &II     &10.3   &8.6    &7.4    &...            &...\\
GY 214   &I      &17.2   &12.0   &7.5    &5/20/00                        &75\\
GY 292   &II     &11.3   &9.3    &7.9    &5/21/00                        &60\\
GY 110\tablenotemark{d}  &II     &10.7   &8.9    &8.0    &...                    &...\\
GSS 28   &II     &9.7    &8.7    &8.1    &5/21/00                        &45\\
GSS 29\tablenotemark{e}  &II     &11.1   &9.2    &8.2    &5/21~\&~22/00  &55~\&~50\\    
GY 6     &I      &13.9   &10.8   &8.3    &5/22/00                        &30\\
GY 308   &II     &11.5   &9.6    &8.3    &5/21/00                        &35\\
GY 182   &I.5    &14.1   &10.6   &7.9    &5/20/00                        &30\\
GY 252   &I.5    &15.2   &11.3   &8.4    &5/20/00                        &35\\
GY 314   &II     &10.8   &9.3    &8.4    &5/21/00                        &40\\
IRS 2    &II     &10.5   &9.1    &8.4    &5/22/00                        &45\\

\enddata
\tablenotetext{a}{Cross reference source names and coordinates are listed 
\citet{barsony1997}}
\tablenotetext{b}{Categorized by LR99} 
\tablenotetext{c}{Photometry from \citet{barsony1997}} 
\tablenotetext{d}{Not observed due to lack of nearby guide stars}
\tablenotetext{e}{Two observations were made of this source}
\end{deluxetable}
\clearpage

\begin{deluxetable}{ccccccccc}
\footnotesize
\tablecaption{Measured stellar parameters of YSOs in $\rho$ Ophiuchi sample\label{tbl-oph.params}}
\tablewidth{5.5in}
\tablehead{
\colhead{Source} & \colhead{T$_{\rm eff}$\tablenotemark{a}} & \colhead{Sp\tablenotemark{b}} &\colhead{$v\sin i$} &\colhead{r$_{\rm K}$} &\colhead{v$_{\rm LSR}$}&\colhead{A$_{\rm K}$\tablenotemark{c}} &\colhead{L$_{\rm bol}$\tablenotemark{d}} & \colhead{$\log g$\tablenotemark{e}}\\
\colhead{}& \colhead{(K)} & \colhead{}&\colhead{(km s$^{-1})$} &\colhead{} &\colhead{(km s$^{-1})$} &\colhead{} &\colhead{(L/L$_{\odot}$)} & \colhead{(cm s$^{-2}$)}} 
\startdata

GY 319   &3410   &M3     &26     &0.9    &3.8 &0.14 &1.20           &3.5\\
GY 314   &3435   &M3     &22     &1.1    &4.0 &0.83 &0.69           &3.5\tablenotemark{f}\\
GY 23    &3490   &M2     &20     &1.2    &0.1 &1.36 &3.03           &3.5\\
GY 292   &3500   &M2     &28     &4.0    &2.8 &1.24 &0.66           &3.5\\
EL 24    &3520   &M2     &21     &4.5    &3.8 &1.16 &1.69           &3.5\\
SR 4     &3585   &M1.5   &14     &3.7    &5.7 &0.35 &0.63           &3.5\\
GSS 28   &3610   &M1.5   &17     &1.2    &4.2 &0.34 &0.63           &3.5\\
GY 308   &3715   &M1     &17     &1.6    &4.1 &1.15 &0.95           &3.5\\
GY 20A   &4460   &K5     &39     &1.6    &6.4 &0.83 &5.19           &3.5\\
IRS 2    &4640   &K3     &12     &0.3    &4.2 &0.67 &1.91           &4.3\tablenotemark{f}\\

\enddata
\tablenotetext{a}{Errors are $\pm$140K (DJ03)}
\tablenotetext{b}{Based on T$_{\rm eff}$ and adopting the spectral 
type--T$_{\rm eff}$ relation of \citet{dejager1987}}
\tablenotetext{c}{Determined using J and H photometry of \citet{barsony1997}
our measured effective temperatures, 
 the extinction law of \citet{martin1990}, and the spectral type--color relation from table A5 of \citet{kenyon1995}}
\tablenotetext{d}{Determined from the observed K--band flux of \citet{barsony1997}, corrected for extinction, and the presence of continuum veiling that we derive at K}
\tablenotetext{e}{Assumed $\log g$ = 3.5 unless otherwise noted}
\tablenotetext{f}{Log~g values measured using spectroscopy}
\end{deluxetable}
\clearpage

\begin{deluxetable}{ccccc}
\footnotesize
\tablecaption{Masses and Ages of $\rho$ Oph sources with spectroscopically measured gravities\label{tbl-mass.age}}
\tablewidth{5.5in}
\tablehead{
\colhead{Source} & \colhead{YSO Evolution} & \colhead{YSO Mass} & \colhead{YSO Age} & \colhead{Luminosity}\\
   & \colhead{Model} & \colhead{(M/M$_{\odot}$)} & \colhead{(Myr)} & \colhead{(L/L$_{\odot}$)}\\
}
\startdata
GY 314 & BCAH\tablenotemark{a} & 0.5  & 1.0  & 0.5\\
(T$_{\rm eff}$=3435K) & SDF\tablenotemark{b}  & 0.34 & 1.5  & 0.3\\
 (log g=3.5)     & PS99\tablenotemark{c} & 0.25 & 1.0  & 0.24\\
      & DM98\tablenotemark{d} & 0.25 & 10   & 0.24\\
\\
IRS 2  & BCAH\tablenotemark{a} & 1.0  & 20   & 0.6\\
(T$_{\rm eff}$=4640K) & SDF\tablenotemark{b}  & 1.2  & 15   & 0.6\\
(log g=4.3)      & PS99\tablenotemark{c} & 1.0  & 20   & 0.5\\
      & DM98\tablenotemark{d} & $>$0.9 & $\sim$70  & $>$0.5\\
\enddata
\tablenotetext{a}{\citet*{baraffe1998}}
\tablenotetext{b}{\citet*{siess2000}}
\tablenotetext{c}{\citet{palla1999}}
\tablenotetext{d}{\citet{dantona1997}}

\end{deluxetable}


\begin{thebibliography}{}
\bibitem[Ali et al.(1995)] {ali1995} Ali, B., Carr, J.S., DePoy, D.L., Frogel, J.A., \& Sellgren, K. 1995, \aj, 110, 2415
\bibitem[Balachandran \& Carr(1994)]{balachandran1994} Balachandran, S. \& Carr, J. 1994, in 8th Cambridge Workshop on Cool Stars, Stellar Systems and the Sun, Caillaut, J.P. ed. ASP Conf. Ser. 64, 264
\bibitem[Baldwin et al.(1973)Baldwin, Frogel, \& Persson]{baldwin1973} Baldwin, J.R., Frogel, J.A., \& Persson, S.E. 1973, \apj, 184, 427
\bibitem[Baraffe et al.(1998)Baraffe, Chabrier, Allard, \& Hauschildt]{baraffe1998} Baraffe, I., Chabrier, G., Allard, F., \& Hauschildt, P.~H.\ 1998, \aap, 337, 403
\bibitem[Baraffe et al.(2002)Baraffe, Chabrier, Allard, \& Hauschildt]{baraffe2002} Baraffe, I., Chabrier, G., Allard, F., \& Hauschildt, P.~H.\ 2002, \aap, 382, 563
\bibitem[Barsony et al.(1997)Barsony, Kenyon, Lada, \& Teuben]{barsony1997} Barsony, M., Kenyon, S.J., Lada, E.A., \& Teuben P.J. 1997, \apjs, 112, 109.
\bibitem[Basri \& Batalha(1990)]{basri1990} Basri, G.~\& Batalha, C.\ 1990, \apj, 363, 654 
\bibitem[Bontemps et al.(2001)]{bontemps2001} Bontemps, S.~et al.\ 2001, \aap, 372, 173
\bibitem[Bouvier et al.(1986)Bouvier, Bertout, Benz, \& Mayor]{bouvier1986} Bouvier, J., Bertout, C., Benz, W., \& Mayor, M.\ 1986, \aap, 165, 110
\bibitem[Calvet et al.(1991)Calvet, Patino, Magris, \& D'Alessio]{calvet1991} Calvet, N., Patino, A., Magris, G.~C., \& D'Alessio, P.\ 1991, \apj, 380, 617
\bibitem[Calvet, Hartmann, \& Strom(1997)]{calvet1997} Calvet, N., Hartmann, L., \& Strom, S.~E.\ 1997, \apj, 481, 912 
\bibitem[Carr \& Tokunaga(1992)]{carr1992} Carr, J.~S.~\& Tokunaga, A.~T.\ 1992, \apjl, 393, L67
\bibitem[Carr et al.(1993)]{carr1993} Carr, J.~S., Tokunaga, A.~T., Najita, J., Shu, F.~H., \& Glassgold, A.~E.\ 1993, \apjl, 411, L37 
\bibitem[Casali \& Eiroa(1996)]{casali1996} Casali, M.~M.~\& Eiroa, C.\ 1996, \aap, 306, 427
\bibitem[Choi \& Herbst(1996)]{choi1996} Choi, P.~I.~\& Herbst, W.\ 1996, \aj, 111, 283
\bibitem[Comer\'{o}n et al.(1993)Comer\'{o}n, Rieke, Burrows, \& Rieke]{comeron1993} Comer\'{o}n, F., Rieke, G.~H., Burrows, A., \& Rieke, M.~J.\ 1993, \apj, 416, 185
\bibitem[de Zeeuw et al.(1999)]{dezeeuw1999} de Zeeuw, P.~T., Hoogerwerf, R., de Bruijne, J.~H.~J., Brown, A.~G.~A., \& Blaauw, A.\ 1999, \aj, 117, 354
\bibitem[D'Antona \& Mazzitelli(1997)]{dantona1997} D'Antona, F.~\& Mazzitelli, I.\ 1997, Mem. Soc. Astron. Italiana, 68, 823
\bibitem[Dent \& Geballe(1991)]{dent1991} Dent, W.~R.~F.~\& Geballe, T.~R.\ 1991, \aap, 252, 775
\bibitem[de Jager \& Nieuwenhuijzen(1987)]{dejager1987} de Jager, C.~\& Nieuwenhuijzen, H.\ 1987, \aap, 177, 217
\bibitem[Doppmann \& Jaffe(2003)]{doppmann2002a} Doppmann, G.~W. \& Jaffe, D.~T. 2003, \aj, submitted (DJ03)
\bibitem[Duquennoy \& Mayor(1991)]{duquennoy1991} Duquennoy, A.~\& Mayor, M.\ 1991, \aap, 248, 485
\bibitem[Edwards et al.(1993)]{edwards1993} Edwards, S.~et al.\ 1993, \aj, 106, 372
\bibitem[Eiroa et al.(2002)]{eiroa2002} Eiroa, C.~et al.\ 2002, \aap, 384, 1038
\bibitem[Elias(1978a)]{elias1978a} Elias, J.~H.\ 1978, \apj, 224, 453
\bibitem[Elias(1978b)]{elias1978b} Elias, J.~H.\ 1978, \apj, 224, 857
\bibitem[Encrenaz(1974)]{encrenaz1974} Encrenaz, P.J.\ 1974, \apj, 189, L135
\bibitem[Evans(1999)]{evans1999} Evans, N.~J.\ 1999, \araa, 37, 311
\bibitem[Fazio et al.(1976)Fazio, Low, Wright, \& Zeilik]{fazio1976} Fazio, G.~G., Low, F.~J., Wright, E.~L., \& Zeilik, M.\ 1976, \apjl, 206, L165
\bibitem[Fernandez \& Eiroa(1996)]{fernandez1996} Fernandez, M.~\& Eiroa, C.\ 1996, \aap, 310, 143 
\bibitem[Folha \& Emerson(1999)]{folha1999} Folha, D.~F.~M.~\& Emerson, J.~P.\ 1999, \aap, 352, 517
\bibitem[Ghez et al.(1993)Ghez, Neugebauer, \& Matthews]{ghez1993} Ghez, A.~M., Neugebauer, G., \& Matthews, K.\ 1993, \aj, 106, 2005
\bibitem[Ghez et al.(1997)Ghez, McCarthy, Patience, \& Beck]{ghez1997} Ghez, A.~M., McCarthy, D.~W., Patience, J.~L., \& Beck, T.~L.\ 1997, \apj, 481, 378
\bibitem[Grasdalen et al.(1973)Grasdalen, Strom, \& Strom]{grasdalen1973} Grasdalen, G.~L., Strom, K.~M., \& Strom, S.~E.\ 1973, \apjl, 184, L53
\bibitem[Greene~\& Young(1992)]{greene1992} Greene, T.P., \& Young, E.T. 1992, \apj, 395, 516
\bibitem[Greene et al.(1993)Greene, Tokunaga, Toomey, \& Carr]{greene1993} Greene, T.~P., Tokunaga, A.~T., Toomey, D.~W., \& Carr, J.~B.\ 1993, \procspie, 1946, 313
\bibitem[Greene et al.(1994)]{GWAYL1994} Greene, T.~P., Wilking, B.~A., Andre, P., Young, E.~T., \& Lada, C.~J.\ 1994, \apj, 434, 614
\bibitem[Greene \& Meyer(1995)]{greene1995} Greene, T.~P.~\& Meyer, M.~R.\ 1995, \apj, 450, 233
\bibitem[Greene~\& Lada(1996)]{greene1996} Greene, T.~P.~\& Lada, C.~J.\ 1996, \aj, 112, 2184
\bibitem[Greene \& Lada(1997)]{greene1997} Greene, T. P. \& Lada, C. J. 1997, \apj, 114, 2157
\bibitem[Greene \& Lada(2000)]{greene2000} Greene, T. P. \& Lada, C. J. 2000, \apj, 120, 430
\bibitem[Greene \& Lada(2002)]{greene2002} Greene, T.~P.~\& Lada, C.~J.\ 2002, \aj, 124, 2185
\bibitem[Gullbring et al.(2000)Gullbring, Calvet, Muzerolle, \& Hartmann]{gullbring2000} Gullbring, E., Calvet, N., Muzerolle, J., \& Hartmann, L.\ 2000, \apj, 544, 927
\bibitem[Hartigan et al.(1989)]{hartigan1989} Hartigan, P., Hartmann, L., Kenyon, S., Hewett, R., \& Stauffer, J.\ 1989, \apjs, 70, 899 
\bibitem[Hartigan et al.(1991)]{hartigan1991} Hartigan, P., Kenyon, S.~J., Hartmann, L., Strom, S.~E., Edwards, S., Welty, A.~D., \& Stauffer, J.\ 1991, \apj, 382, 617 
\bibitem[Hartigan et al.(1995)Hartigan, Edwards, \& Ghandour]{hartigan1995} Hartigan, P., Edwards, S., \& Ghandour, L.\ 1995, \apj, 452, 736
\bibitem[Hartmann et al.(1997)Hartmann, Cassen, \& Kenyon]{hartmann1997} Hartmann, L., Cassen, P., \& Kenyon, S.~J.\ 1997, \apj, 475, 770
\bibitem[Hartmann(2001)]{hartmann2001} Hartmann, L.\ 2001, \aj, 121, 1030
\bibitem[Hartmann(2002)]{hartmann2002} Hartmann, L.\ 2002, \apjl, 566, L29
\bibitem[Hauschildt et al.(1999)Hauschildt, Allard, \& Baron]{hauschildt1999} Hauschildt P. H., Allard, F. \& Baron, E. 1999, \apj, 512, 377
\bibitem[Herbst et al.(2000)Herbst, Rhode, Hillenbrand, \& Curran]{herbst2000} Herbst, W., Rhode, K.~L., Hillenbrand, L.~A., \& Curran, G.\ 2000, \aj, 119, 261
\bibitem[Herbst et al.(2002)]{herbst2002} Herbst, W., Bailer--Jones, C.~A.~L., Mundt, R., Meisenheimer, K., \& Wackermann, R.\ 2002, \aap, 396, 513
\bibitem[Hinkle et al.(1998)]{hinkle1998} Hinkle, K.~H., Cuberly, R.~W., Gaughan, N.~A., Heynssens, J.~B., Joyce, R.~R., Ridgway, S.~T., Schmitt, P., \& Simmons, J.~E.\ 1998, \procspie, 3354, 810
\bibitem[Johns--Krull et al.(1999) Johns--Krull, Valenti, \& Koresko]{johnskrull1999} Johns--Krull, C.~M., Valenti, J.~A., \& Koresko, C.\ 1999, \apj, 516, 900
\bibitem[Johns--Krull \& Valenti(2001)]{johnskrull2001a} Johns--Krull, C.~M.~\& Valenti, J.~A.\ 2001, \apj, 561, 1060
\bibitem[Johns--Krull et al.(2001)]{johnskrull2001b} Johns--Krull, C.~M., Valenti, J.~A., Piskunov, N.~E., Saar, S.~H., \& Hatzes, A.~P.\ 2001, Magnetic Fields Across the Hertzsprung--Russell Diagram, ASP Conference Proceedings Vol.~248.~Edited by G.~Mathys, S.~K.~Solanki, and D.~T.~Wickramasinghe.~ ISBN: 1-58381-088-9.~ San Francisco: Astronomical Society of the Pacific, 2001., p.527, 527
\bibitem[Kenyon \& Hartmann(1990)]{kenyon1990} Kenyon, S.~J.~\& Hartmann, L.~W.\ 1990, \apj, 349, 197
\bibitem[Kenyon \& Hartmann(1995)]{kenyon1995} Kenyon, S.~J.~\& Hartmann, L.\ 1995, \apjs, 101, 117
\bibitem[Kleinmann \& Hall(1986)]{kleinmann1986} Kleinmann, S. G., \& Hall, D. N. B. 1986, \apjs, 62, 501
\bibitem[K\"{o}nigl(1991)]{konigl1991} K\"{o}nigl, A.\ 1991, \apjl, 370, L39
\bibitem[Lada \& Wilking(1984)]{lada1984} Lada, C.~J.~\& Wilking, B.~A.\ 1984, \apj, 287, 610
\bibitem[Lada et al.(1991)Lada, Evans, Depoy, \& Gatley]{lada1991} Lada, E.~A., Evans, N.~J., Depoy, D.~L., \& Gatley, I.\ 1991, \apj, 371, 171
\bibitem[Lada \& Adams(1992)]{lada1992} Lada, C.~J.~\& Adams, F.~C.\ 1992, \apj, 393, 278
\bibitem[Lada \& Lada(2003)]{lada2003} Lada, C.~J.~\& Lada, E.~A.\ 2003, \araa, in press
\bibitem[Luhman \& Rieke(1999)]{luhman1999} Luhman. K. L. \& Rieke, G. H. 1999, ApJ, 525, 440 (LR99)
\bibitem[Martin \& Whittet(1990)]{martin1990} Martin, P. G. \& Whittet, D.\ C. B. 1990,\apj, 357, 113
\bibitem[Mathieu et al.(2000)Mathieu, Ghez, Jensen, \& Simon]{mathieu2000} Mathieu, R. D., Ghez, A. M., Jensen, E. L. N., \& Simon, M. 2000, in Protostars and Planets IV, ed. V. Mannings, A. P. Boss, \& S. S. Russell (Tucson: Univ. Arizona Press), 703
\bibitem[McLean et al.(1995)]{mclean1998} McLean, I.~S., Becklin, E.~E., Figer, D.~F., Larson, S., Liu, T., \& Graham, J.\ 1995, \procspie, 2475, 350
\bibitem[Meyer et al.(1997)Meyer, Calvet, \& Hillenbrand]{meyer1997} Meyer, M.~R., Calvet, N., \& Hillenbrand, L.~A.\ 1997, \aj, 114, 288
\bibitem[Miller \& Scalo(1978)]{miller1978} Miller, G.~E.~\& Scalo, J.~M.\ 1978, \pasp, 90, 506
\bibitem[Muzerolle et al.(1998)Muzerolle, Calvet, \& Hartmann(1998)]{muzerolle1998a} Muzerolle, J., Calvet, N., \& Hartmann, L.\ 1998, \apj, 492, 743
\bibitem[Muzerolle, Hartmann, \& Calvet(1998)]{muzerolle1998b} Muzerolle, J., Hartmann, L., \& Calvet, N.\ 1998, \aj, 116, 2965
\bibitem[Myers \& Ho (1975)]{myers1975} Myers, P.C., \& Ho, P.T.P.\ 1975, \apj, 202, L25
\bibitem[Najita et al.(1996)]{najita1996} Najita, J., Carr, J.~S., Glassgold, A.~E., Shu, F.~H., \& Tokunaga, A.~T.\ 1996, \apj, 462, 919
\bibitem[Padgett(1996)]{padgett1996} Padgett, D.~L.\ 1996, \apj, 471, 847
\bibitem[Palla \& Stahler(1999)]{palla1999} Palla, F.~\& Stahler, S.~W.\ 1999, \apj, 525, 772
\bibitem[Palla \& Stahler(2000)]{palla2000} Palla, F.~\& Stahler, S.~W.\ 2000, \apj, 540, 255
\bibitem[Ram\'{i}rez et al.(1997)]{ramirez1997} Ram\'{i}rez, S.V., DePoy, D.L., Frogel, J.A., Sellgren, K., \& Blum, R.D. 1997, \aj, 113, 1411
\bibitem[Rothman et al.(1992)]{rothman1992} Rothman, L.~S.~et al.\ 1992, Journal of Quantitative Spectroscopy and Radiative Transfer, 48, 469
\bibitem[Shu et al.(1987)Shu, Adams, \& Lizano]{shu1987} Shu, F.~H., Adams, F.~C., \& Lizano, S.\ 1987, \araa, 25, 23
\bibitem[Shu et al.(1994)]{shu1994} Shu, F., Najita, J., Ostriker, E., Wilkin, F., Ruden, S., \& Lizano, S.\ 1994, \apj, 429, 781
\bibitem[Siess et al.(2000)Siess, Dufour, \& Forestini]{siess2000} Siess, L., Dufour, E., \& Forestini, M. 2000, \aap, 358, 593
\bibitem[Sneden(1973)]{sneden1973} Sneden, C. 1973 PhD Thesis, University of Texas at Austin
\bibitem[Stassun et al.(1999)Stassun, Mathieu, Mazeh, \& Vrba]{stassun1999} Stassun, K.~G., Mathieu, R.~D., Mazeh, T., \& Vrba, F.~J.\ 1999, \aj, 117, 2941
\bibitem[Strom et al.(1988)Strom, Strom, Kenyon, \& Hartmann]{strom1988} Strom, K.~M., Strom, S.~E., Kenyon, S.~J., \& Hartmann, L.\ 1988, \aj, 95, 534
\bibitem[Strom et al.(1995)Strom, Kepner, \& Strom]{strom1995} Strom, K.~M., Kepner, J., \& Strom, S.~E.\ 1995, \apj, 438, 813
\bibitem[Strom \& Strom(1994)]{strom1994} Strom, K.~M.~\& Strom, S.~E.\ 1994, \apj, 424, 237
\bibitem[Thompson(1985)]{thompson1985} Thompson, R.~I.\ 1985, \apjl, 299, L41
\bibitem[Tout, Livio, \& Bonnell(1999)]{tout1999} Tout, C.~A., Livio, M., \& Bonnell, I.~A.\ 1999, \mnras, 310, 360
\bibitem[van Dishoeck \& de Zeeuw(1984)]{vandishoeck1984} van Dishoeck, E.~F.~\& de Zeeuw, T.\ 1984, \mnras, 206, 383
\bibitem[Vrba et al.(1975)Vrba, Strom, Strom, \& Grasdalen]{vrba1975} Vrba, F.~J., Strom, K.~M., Strom, S.~E., \& Grasdalen, G.~L.\ 1975, \apj, 197, 77
\bibitem[Wilking~\& Lada(1983)]{wilking1983} Wilking, B.A., \& Lada, C.J. 1983, ApJ, 274, 698
\bibitem[Wilking et al.(1989)Wilking, Lada, \& Young]{wilking1989} Wilking, B.~A., Lada, C.~J., \& Young, E.~T.\ 1989, \apj, 340, 823

\end{thebibliography}
\end{document}